\documentclass{aa}  
\usepackage{graphicx}
\usepackage{txfonts}
\usepackage{hyperref}
\hypersetup{
    colorlinks=true,
    linkcolor=blue,
    filecolor=magenta,      
    urlcolor=cyan,
    citecolor=blue
    }
\usepackage[section]{placeins}

\newcommand{\psls}{\texttt{PSLS}}
\newcommand{\pyspot}{\texttt{pyspot}}
\newcommand{\pytransit}{\texttt{PyTransit}}
\newcommand{\wotan}{\texttt{W\~otan}}
\newcommand{\tls}{\texttt{TLS}}
\newcommand{\cetra}{\texttt{CETRA}}
\newcommand{\nuance}{\texttt{nuance}}
\newcommand{\umbra}{\texttt{umbra}}

\begin{document} 

\title{Detecting transiting exoplanets in simulated PLATO data}
\subtitle{A comparison of light curve filter and transit search algorithms}
\titlerunning{Filter and search comparison}

\date{Received 14 April 2026 / Accepted 11 September 2026}

\author{
Geert Jan Talens\inst{1} \thanks{Corresponding author: \email{phys2523@ox.ac.uk}} \and
Leigh C. Smith\inst{2} \and
Luca Malavolta \inst{3,4} \and 
Lionel Garcia\inst{5} \and
Simon Hodgkin \inst{2} \and
Suzanne Aigrain\inst{1}
}

\institute{
Denys Wilkinson Building, Department of Physics, University of Oxford, OX1 3RH, United Kingdom 
\and 
Institute of Astronomy, University of Cambridge, Madingley Rd, Cambridge CB3 0HA, United Kingdom \and
Dipartimento di Fisica e Astronomia “Galileo Galilei”, Università di Padova, Vicolo dell’Osservatorio 3, 35122 Padova, Italy
\and
INAF - Osservatorio Astronomico di Padova, Vicolo dell'Osservatorio 5, 35122 Padova, Italy
\and
Center for Computational Astrophysics, Flatiron Institute, New York, NY, USA 
}

\abstract
{}
{Our goal is to test and compare transit search methodologies, and provide recommendations as to the best practices to be implemented by the upcoming ESA PLATO mission, in order to achieve its goal of detecting Earth-like planets in the habitable zones of Sun-like stars.}
{We generate simulated two-year PLATO light curves with injected planets with sizes 0.5--2.0 $R_\oplus$, and compare the performance of the {\cetra}, {\umbra}, and {\nuance} transit search methods. {\cetra} and {\umbra} require pre-filtered light curves and we test the biweight, Huber spline, Lowess, and YSD-Lowess filters. {\nuance} uses a Gaussian process to model the stellar variability at all proposed transit parameters.}
{{\nuance} achieves the best performance, recovering 61.2\% signals using a simple harmonic oscillator Gaussian process kernel, and it performs especially well for both hot and fast-rotating stars. However, when multiple Huber spline filter windows are considered, both {\cetra} and {\umbra} can match {\nuance}'s performance. {\cetra} matches {\nuance} with four windows, recovering 61.3\% of signals. {\umbra} matches {\nuance} with two windows, recovering 61.7\% of signals. Recovery rates exceeding {\nuance} are possible by using more windows. The computational cost of running {\nuance} is so high that running {\cetra} or {\umbra} on multiple windows is preferred.}
{The PLATO pipeline should use the Huber spline with a range of window sizes as its primary light curve filter. {\cetra} should be updated to use the warped least-squares templates implemented in {\umbra}. The use of {\nuance} should be considered for hot, rapidly rotating stars.}

\keywords{planets and satellites: detection -- methods: numerical -- methods: data analysis -- techniques: photometric}

\maketitle
\nolinenumbers

\section{Introduction}
\label{sec:intro}

The detection of transit events in photometric time-series is today the most successful method for discovering and characterizing planets outside our own solar system. Photometric time-series are comparatively easy to obtain in bulk observationally, and transiting exoplanets give access not only to the planetary mass and radius, but also to atmospheric properties through transmission and emission spectroscopy and phase curve observations. However, though thousands of transiting exoplanets have been detected and validated by ground- and space-based surveys \citep{https://doi.org/10.26133/nea37}, and hundreds of planets have been characterized in various levels of detail, no true Earth analog planet has of yet been found. 

The upcoming PLAnetary Transits and Oscillations of stars (PLATO) mission \citep{PLATO-2025} aims to find Earth-like planets in the habitable zone (HZ) of their host stars, by monitoring the brightness of main sequence and subgiant stars with spectral types F5--K7. The nominal mission duration is four years, with at least the first two years dedicated to observing the long-duration observation phase south 2 (LOPS2) field \citep{Nascimbeni-2025}. The resulting time-series photometry will have the stellar variability removed before being searched for transiting signals using the Cambridge Exoplanet Transit Recovery Algorithm ({\cetra}, \citealt{Smith-2025}). 

The removal of stellar variability signals prior to transit searches has been performed with many different types of filters, in both the time and frequency domain. A summary of available time domain filters was presented by \citet{Hippke-2019-Wotan}, and these filters are implemented in the {\wotan} package. Wavelets \citep{Jenkins-2002} are the most common frequency domain filter and were adopted by the \emph{Kepler}, \emph{K2}, and \emph{TESS} missions \citep{Jenkins-2020}. Though some filters are preferred for transit detection, the optimal filter for a specific dataset depends on the instrument, the stellar sample observed, and on the transit durations of interest. Many approaches exist for detecting exoplanet signals in time-series photometry. The most successfully of which is box least-squares \citep[BLS;][]{Kovacs-2002}, and its more recent successor transit least-squares (TLS) published by \citet{Hippke-2019-TLS} and implemented in several python packages, including {\tls} and {\cetra}. Throughout this paper we use TLS (normal font) to refer to the algorithm, and {\tls} (typewriter font) to refer to the software package. 

In this paper we test the performance of different filter and search algorithms on simulated two-year PLATO light curves, building on work done by \citet{Canocchi-2023}. We benchmark {\cetra}'s performance on light curves filtered with the biweight, Huber spline, Lowess and Lowess based young star detrending (YSD-Lowess) filters against the {\umbra} and {\nuance} transit search codes. The {\umbra} code (Talens et al., in prep) uses transit templates that account for the effect of traditional variability filters on the transit shape. {\nuance} \citep{Garcia-2024} uses a Gaussian process (GP) to simultaneously model the transit and stellar variability.

Our analysis is described in Sect.~\ref{sec:analysis}, with Sect.~\ref{sec:simlcs} describing the simulated PLATO light curves, Sect.~\ref{sec:filters} defining the variability filters tested, and Sect.~\ref{sec:search} presenting the transit search methodologies explored. Section~\ref{sec:results} presents the results of the comparison, and we summarize our finding and make recommendations in Sect.~\ref{sec:conclusions}.

\section{Analysis}
\label{sec:analysis}

Our analysis starts with the generation of a set of simulated PLATO light curves. Next, one of four variability filters is used to remove nuisance signals and systematics from the light curves. Finally, three different transit search methods are used to try and recover the injected signals.

\begin{figure}[tbp!]
    \centering
    \includegraphics[width=1\linewidth]{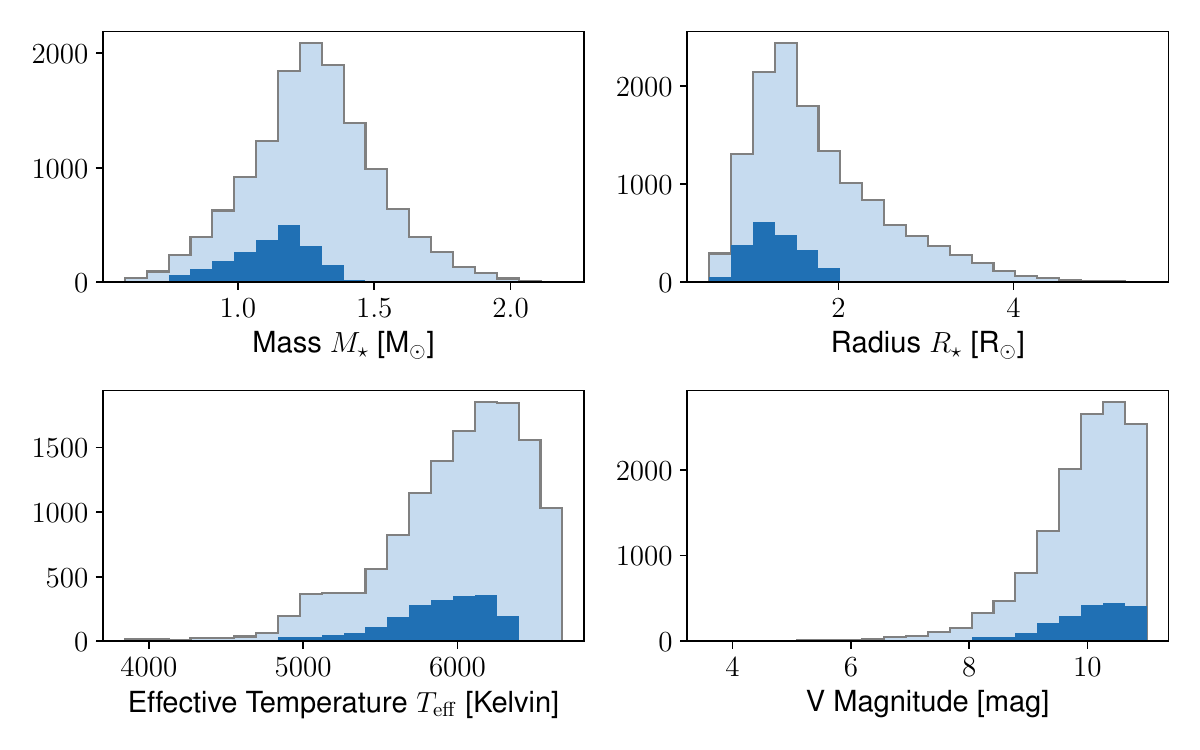}
    \caption{Distribution of the stellar parameters. Light blue: the PLATO P1 + P2 sample. Dark blue: the 2000 stars randomly selected after cuts on $\log g$, $T_{\rm{eff}}$, and $V$ were applied.}
    \label{fig:stellar_sample}
\end{figure}

\subsection{Simulated light curves}
\label{sec:simlcs}

In order to test the different light curve filter and transit search algorithms, we need a suitable simulated dataset. A number of simulated datasets have been produced within the PLATO Mission Consortium, but none are suitable for our purposes. The simulations presented in \citet{Jannsen-2024} were generated with somewhat pessimistic assumptions regarding instrumental systematics; the simulations of \citet{Breton-2024} have a cadence of two hours which is not suitable for transit detection; and the simulations of \citet{Jannsen-2025} focused on stellar oscillations. Furthermore, we now have a better understanding of some important parameters such as the duration and frequency of data gaps both within and between the quarters. We therefore decided to simulate a new, dedicated dataset for this study. 

The main steps in the process were as follows. First, we used the latest version of the PLATO input catalogue (PIC) for the LOPS2 field \citep{Nascimbeni-2025} to select stars with representative parameters. We then simulated light curves including photon noise, systematic trends, granulation and oscillations using the PLATO Solar-like Light curve Simulator\footnote{\url{https://sites.lesia.obspm.fr/psls/}} \citep[{\psls},][]{Samadi-2019}. Starspot signals were generated separately using an updated version of the {\pyspot} package \citep{Talens-2025-pyspot}. Finally, planetary transits were simulated using {\pytransit} \citep{Parviainen-2015}.

\subsubsection{Stellar parameters}
\label{sec:simlcs-stars}

The LOPS2 field, located in the southern hemisphere, will be the first long-duration pointing field observed by PLATO after it launches in early 2027, and will be observed almost continuously for a total of two years. We took the stellar parameters for our simulations from the PIC for this field \citep{Nascimbeni-2025}. We focussed on the bright stellar samples, known as P1 and P2, which have $V < 11$~mag. We selected these stars from the PIC (version 2.1.0.1), resulting in  a total of 13,330 stars. We further selected stars with $\log g > 4$ to exclude giant stars, and restricted ourselves to stars fainter than $V=8$~mag, as stars brighter than this limit may be saturated and cannot be simulated with {\psls}, and with effective temperature in the range $4594 \leq T_{\rm{eff}} \leq 6334$~K, as this is the range for which the scaling laws used to simulate star-spot signals with the \pyspot\ code are applicable. These cuts reduced the sample to 6552 stars, from which we randomly selected 2000 to be simulated. The stellar parameters of the selected stars are compared to the full P1 $+$ P2 sample in Fig.~\ref{fig:stellar_sample}.

We randomly assigned a stellar inclination ($i_\star$) between 0 and 90 degrees. The stellar inclination affects the star-spot signal and also the non-radial modes of the oscillation signal generated by {\psls}. The orbital inclination of the planets we injected later is drawn independently of the stellar inclination. We further assumed a solar metallicity ($[\mathrm{M/H}] = 0$~dex) for all stars. Metallicity only enters the simulations via the limb-darkening parameters used to simulate the transits, which we computed from the 4-parameter limb-darkening coefficients for the PLATO bandpass presented in \citet{Kostogryz-2022}. 

\subsubsection{Transit parameters}
\label{sec:simlcs-planets}

\begin{figure}[tbp!]
    \centering
    \includegraphics[width=1\columnwidth]{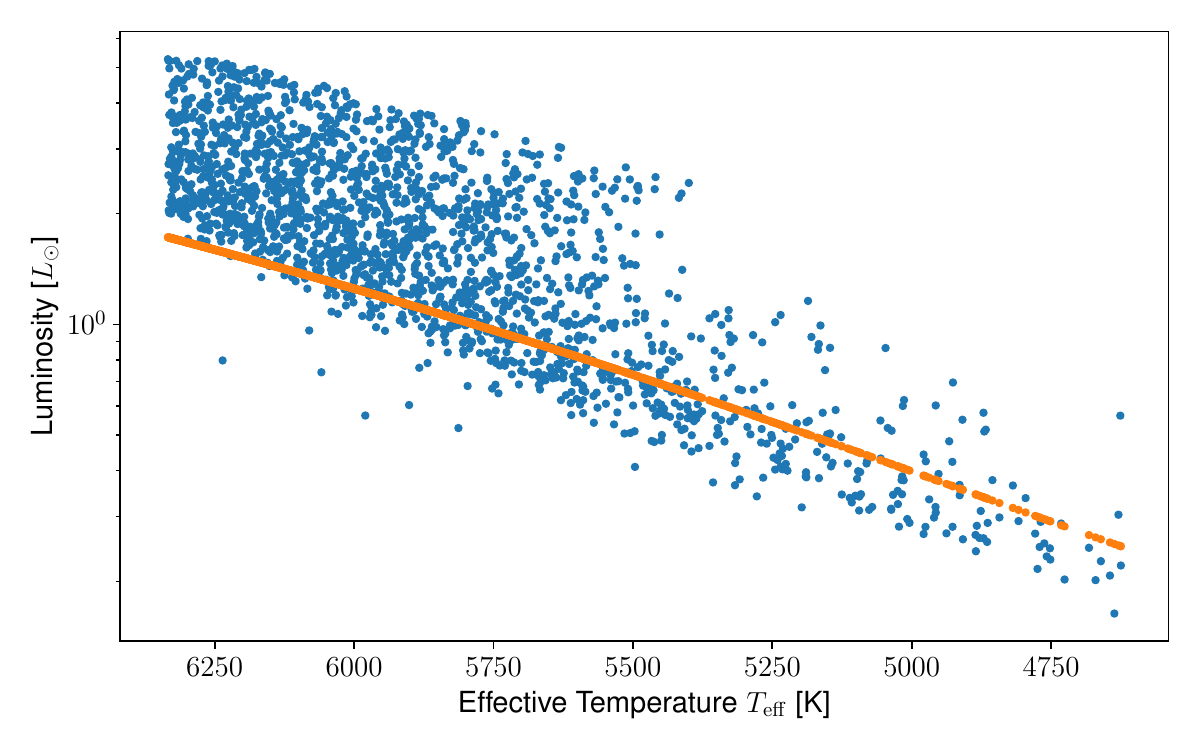}
    \caption{The PIC stellar luminosities (blue), and the main sequence equivalent luminosities (orange) for the 2000 simulated stars.}
    \label{fig:luminosity}
\end{figure}

\begin{figure*}[tbp!]
    \centering
    \includegraphics[width=1\columnwidth]{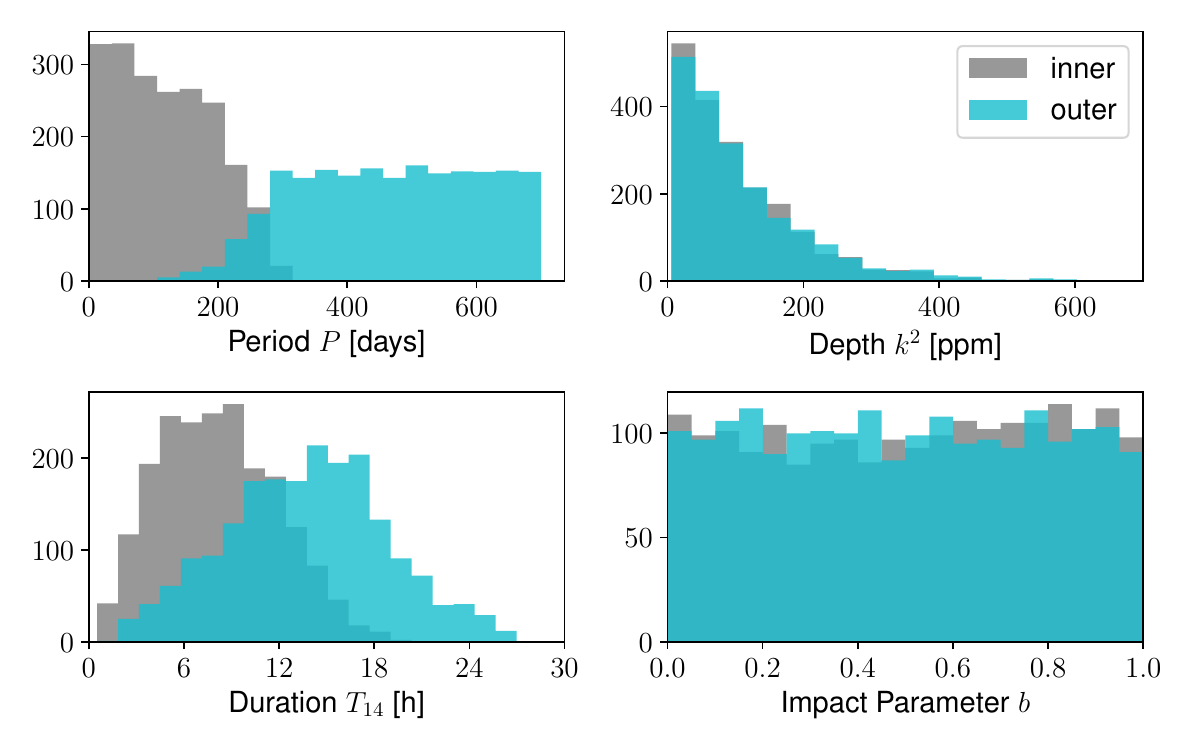}
    \includegraphics[width=1\columnwidth]{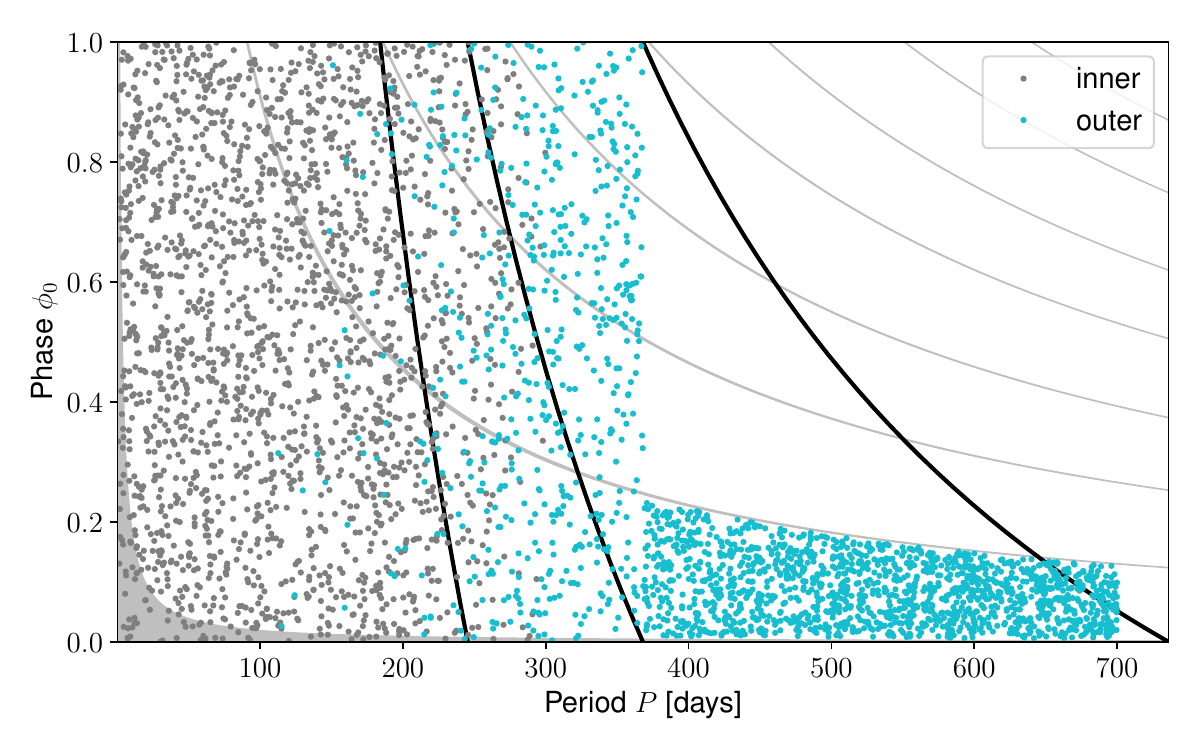}
    \caption{Transit parameters for the inner (grey) and outer (cyan) transit samples. Left: Distribution of the transit parameters in the inner and outer samples. Right: Orbital phase as a function of orbital period for the transit samples. Grey bands indicate the primary locations of the two~day gaps caused by PLATO's quarterly rolls. Solid black lines indicate from right to left the thresholds for two, three and four observed transits in the two year baseline. Thresholds for higher numbers of transits are omitted for clarity.}
    \label{fig:planet_sample}
\end{figure*}

For each of the 2000 stars, we generated two sets of transit parameters: one set with shorter orbital periods, dominated by systems which transit at least three times during a two year pointing; and one set with longer orbital periods, dominated by systems generating two transits during a two year pointing. We refer to these as the ``inner'' and ``outer'' transit samples.  

For each star the allowed orbital periods in the outer samples were computed from modified versions of the recent Venus and early Mars HZ boundaries presented by \citet{Kopparapu-2013}. Without modification the recent Venus and early Mars boundaries would result in outer planets with orbital periods significantly longer than the two year duration of the LOPS2 pointing (up to six years in some cases). This is because a significant fraction of the stars in the P1 sample are subgiants, and using such periods would have resulted in an outer sample dominated by mono-transits. Since the purpose of this work is primarily to test transit searches that rely on the repeating nature of such signals in photometric time-series, we deemed this undesirable. Instead we replaced the PIC luminosity with the ``main sequence equivalent'' luminosity when computing HZ boundaries for each star. The main sequence equivalent luminosity is the luminosity a star of a given effective temperature would have it it was on the main sequence, as illustrated in Fig.~\ref{fig:luminosity}. Using the main sequence equivalent luminosity ensures that the modified HZ boundaries depends only on the stellar effective temperature when expressed in terms of the semi-major axis. The inner simulations used orbital periods interior to the modified HZ, while the outer simulations used orbital periods that placed them in the modified HZ for each star. We want to emphasize that the outer sample is a HZ inspired stress-test sample and our results for this sample have no bearing on PLATO yields for the actual habitable zone.

The maximum orbital period was further constrained to $P < 700$~d. When $P > 368$~d mono-transits can occur, to prevent this we constrained the orbital phases so the first transit is fully observed in the 1st quarter of observations. This choice biases this part of parameter space toward duo- rather than mono-transits, ensuring that periodic search algorithms can be tested on the simulated light curves. Overall, these choices mean that the outer sample is biased in favour of systems that are easier to detect (shorter periods and more favourable epochs) than a sample with completely uniform epochs and periods across the modified HZ range.

For the remaining transit parameters we selected impact parameters uniform between 0 and 1; planet radii uniform between $0.5~R_\oplus$ and $2.0~R_\oplus$, and assumed all orbits are circular. Figure~\ref{fig:planet_sample} shows the final transit parameters for the inner and outer samples, and further illustrates the constraint on orbital period and phase. A figure showing the correlations between the transit and stellar parameters in the inner and outer samples is available in  Appendix~\ref{app:supplementary-figures}. 

\subsubsection{Model evaluation}
\label{sec:model-eval}

We used version 1.8 of {\psls} \citep{Samadi-2019} to generate the observation times and simulate the effects of photon noise, long-term systematic trends, granulation and oscillations. Unlike \texttt{PlatoSim} \citep{Jannsen-2024}, which injects astrophysical signals at the pixel level, then simulates photon noise and instrumental systematic effects, {\psls} uses precomputed scaling laws to simulate the noise and stellar component separately, based on a star's intrinsic parameters (such as $T_{\rm eff}$, $\log g$) as well as its magnitude and accounting for representative drift in the position on the detector. Using precomputed scaling laws makes {\psls} much cheaper in terms of computing time and memory, albeit potentially less accurate. 

For the observation times, we configured {\psls} to simulate individual quarters lasting 90, 95, 89, 97, 84, 95, 89, and 97~days respectively, resulting in a total pointing of 736~days. These quarter durations reflect the expectation that the exact duration of the PLATO quarters will vary by a few days from one quarter to the next. We also included the effects of data gaps from quarterly rolls and subsequent calibration observations, excluding two~days at the start of each quarter, and 20~minute gaps due to spacecraft momentum dumps occurring every $72\pm2$ hours. These numbers are all based on current best engineering estimates that might be reviewed during commissioning.

For some stars the long-term systematic trend produced by {\psls} may have a peak-to-peak amplitude exceeding $5000$~ppm in one or more quarters, which we consider unrealistically large. If the peak-to-peak amplitude exceeded this threshold, we applied a re-scaling to the relevant quarter to achieve a lower peak-to-peak amplitude drawn from a truncated normal distribution with parameters $\mu=1250~\rm{ppm},~\sigma=400~\rm{ppm}$ and truncated to the range 350--2500~ppm. This distribution was empirically determined from the long-term systematics with peak-to-peak amplitudes below $5000$~ppm. The multi-day time-scale of the systematics component is sufficiently long that this re-scaling is unlikely to impact the work presented here. 

An updated version of {\pyspot}\footnote{\url{https://github.com/talensgj/pyspot}} \citep{Talens-2025-pyspot} was used to generate the signal produced by star-spot emerging and decaying on the rotating stellar disk. The spot signal is primarily characterized by the rotation period of the star ($P_{\rm{rot}}$) and the cycle amplitude ($A_{\rm{cyc}}$). Both $P_{\rm{rot}}$ and $A_{\rm{cyc}}$ are set by {\pyspot} based on the prescriptions of \citet{Meunier-2019}. Our new version of {\pyspot} uses the \citet{Kipping-2012} spot model to include arbitrary limb-darkening laws; implements squared-exponential evolution for the spot size to avoid discontinuities in the derivative; and fixes several bugs in the way the level of stellar activity was calculated from the provided stellar parameters. The spots themselves were assumed to be fully dark. The spot model was evaluated at 5-min cadence and interpolated to the time-stamps provided by {\psls} using a cubic spline. We only considered the effect of spots on the overall brightness of the star, and did not explicitly model spot-crossings. Instead, the variability due to spots and the transit signal were multiplied when obtaining the final light curve.

Finally, transit models were evaluated using the {\tt RoadRunnerModel} from {\pytransit} \citep{Parviainen-2015}. To ensure sufficient precision in the case of shallow transits, we used parameters {\tt nzin=40}, {\tt nzlimb=60}, and {\tt ng=250}. 

\subsubsection{Combining signals}
\label{sec:simlcs-final}

\begin{figure*}[tbp!]
    \centering
    \includegraphics[width=1\linewidth]{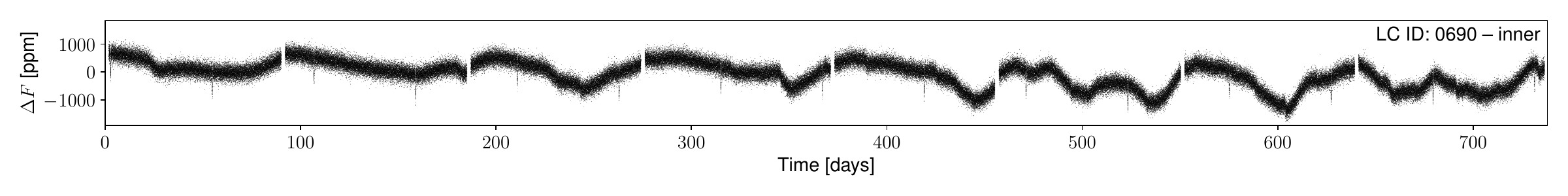}
    \caption{Simulated 25~s cadence PLATO light curve. This light curve shows moderate rotational modulations due to spots and contains a $\sim530$~ppm transit signal with $P\sim52$~d. For clarity, only every 10th observation is shown.}
    \label{fig:example_lc}
\end{figure*}

We stored all components of our simulations separately, allowing us to create variations on the light curves that exclude certain signals. These master files are available from the Oxford research archive\footnote{\url{https://ora.ox.ac.uk/objects/uuid:682bcbb6-ed68-42ca-8c00-277a3aea64bd}} \citep{Talens-2026a}. For the analysis presented in this paper, we generated several variations of the 2000 light curves. First we generated light curves which contain all stellar and instrumental components, i.e. the effects of photon noise, systematics, granulation, oscillations, and spots, but with variants for the inner and outer transit samples as well as a planet-free sample. We refer to the combined inner and outer samples as the ``joint'' sample.

We also generated versions of the light curves, which do not contain the effects of systematics and spots, and which therefore do not require filtering to detect the transits. We refer to these as the ``simple'' light curves, and used them to benchmark the algorithms' performance in the absence of variability. For the purposes of this study we did not consider multi-planet systems or astrophysical false positives such as eclipsing binaries. Figure~\ref{fig:example_lc} shows one of the inner light curves with the effects of systematics, spots, and transits clearly visible. 

\subsection{Light curve filters}
\label{sec:filters}

The majority of transit search codes tested in this work (see Sec.~\ref{sec:search}) require pre-filtering of the light curves to remove stellar variability in order to be able to detect transit signals. The advantage of pre-filtering light curves in this way (compared to modelling the variability simultaneously with the putative transit signal) is that it is typically fast, as the filter is not recomputed for each trial transit model. However, filtering necessarily alters the transit signal, and the filter window must be tuned carefully to  remove as much of the stellar variability and preserve as much of the transit signal as possible. If the window is too wide, stellar variability will not be sufficiently filtered, reducing the detectability of the transits. However, if the window is too short, the transits themselves will be distorted or removed entirely. As transits of planets on wider orbits have longer durations, the timescales of stellar variability and long-period transits tend to overlap, making it more challenging to find a balance between these requirements. 

\citet{Canocchi-2023} previously tested a number of filtering algorithms on simulated PLATO data, using the {\tls} package \citep{Hippke-2019-TLS} for transit detection. They tested a number of general-purpose filters for time-series data, as well as more specialized filters designed specifically for transit searches. They concluded that the best general filters were the biweight and Huber spline \citep{Hippke-2019-Wotan} and the best specialized filters were Notch and LOCoR (locally optimized combination of rotations) \citep{Rizzuto-2017} and YSD-Lowess \citep{Battley-2020}. Overall \citet{Canocchi-2023} prefer the Notch and LOCoR filter, depending on the length of the time-series and the size of the injected planets. However, their work focused primarily on detections of short orbital periods ($P < 40$~d) in single quarters of data taken from a set of 100 two year simulated PLATO stellar light curves injected with different transit signals. This study expands on their work in a number of ways. We use 2000 stellar light curves injected with typically much longer orbits and focusing on transits of Earth-size planets. We always use the full two year light curves, and we consider not just different light curve filters across a wider range of filter windows, but also different transit search algorithms. 

In this work, we consider four filters, for use with the {\cetra} and {\umbra} transit search algorithms: the biweight, Huber spline, and Lowess methods from the {\wotan} package and a custom YSD-Lowess implementation. We do not consider the Notch and LOCoR filter considered best by \citet{Canocchi-2023} as the computational expense of this filter was deemed too great for inclusion in the PLATO pipeline. Each of the tested filters is described in more detail below. In some figures we will also show light curves filtered with a moving mean (boxcar filter), but since the mean is not robust against outliers we do not perform transit searches on the mean-filtered data.
Unless otherwise noted, we applied the filters to individual quarters of the simulated PLATO light curves, as both the two~day quarter gap and changing systematics prevent continuity between the quarters. 

\begin{figure*}[tbp!]
    \centering
    \includegraphics[width=1\linewidth]{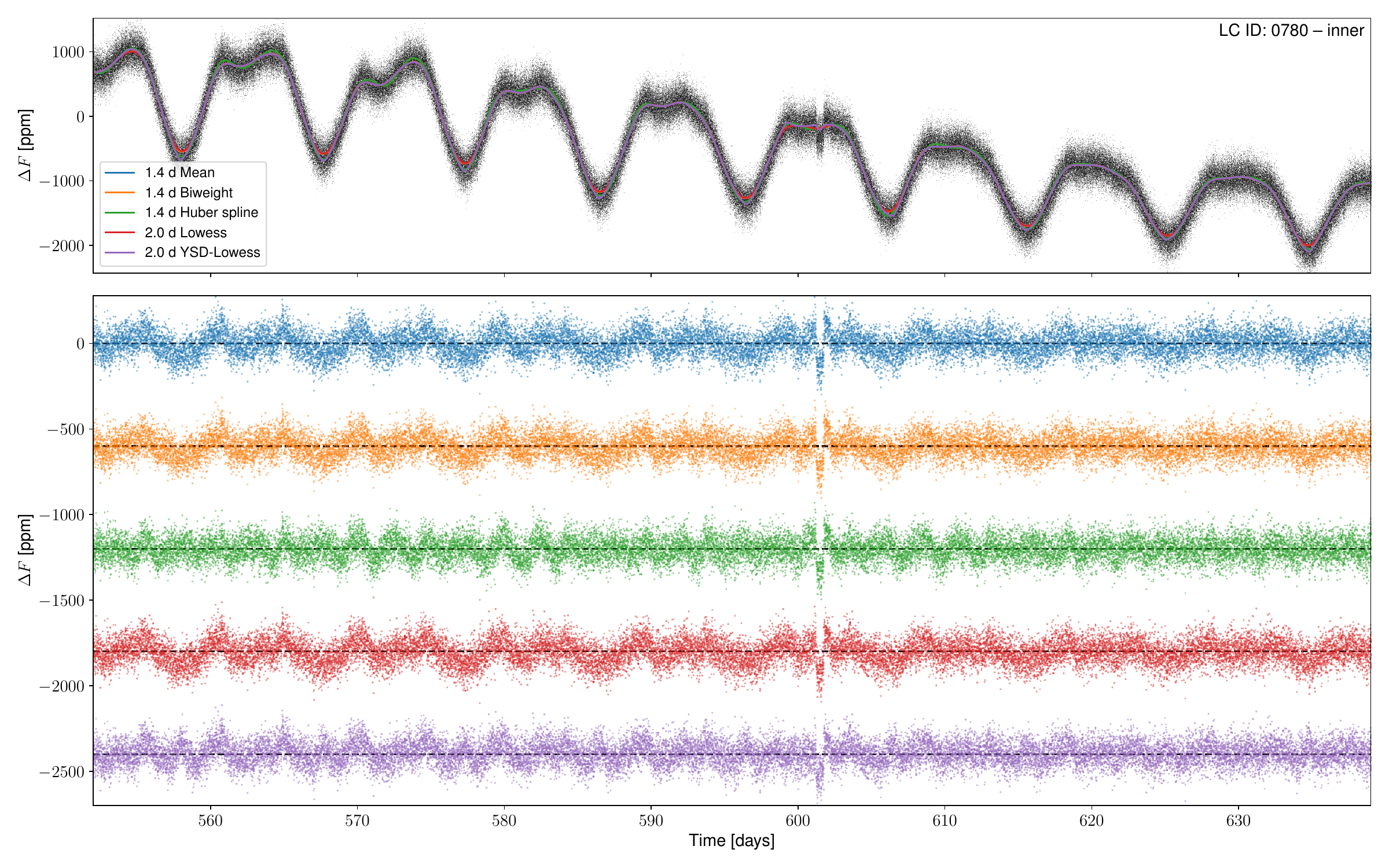}
    \caption{Effect of light curve filters on a single quarter segment of a moderately active star. Top: single quarter segment of a simulated PLATO light curve at 25~s cadence (black dots), and trends obtained by running the mean, biweight, Huber spline, Lowess, and YSD-Lowess light curve filters. Bottom: the light curve after dividing by each filter's trend and binning to 5~min cadence. The light curves obtained with different filters are offset, with the baseline for each indicated by the dashed lines.}
    \label{fig:filters_quarter}
\end{figure*}

\paragraph{Biweight.} Tukey's biweight \citep{Mostellor-1977} is a robust estimator, which iteratively computes a robust estimate of the mean of the data within a window of width $w$ around each data point, by weighing points according to their distance from the trend. In the context of transit detection, if the window $w$ is much longer than the transit duration $T_{14}$, and individual transits have high enough signal-to-noise ratio (S/N), the biweight filter largely avoids modifying the transit shape, but these conditions are not always met in practice. Many stars show variability on time-scales such that $w \sim T_{14}$ would be needed to remove it. Furthermore, the individual transits of Earth-sized planets around Sun-like stars are unlikely to have high S/N. Nevertheless, a robust estimator like the biweight is desirable, as the PLATO photometry is expected to contain outliers of both astrophysical and instrumental origins, which should be excluded when filtering the light curve.

\paragraph{Huber spline.} The Huber spline filter as implemented in {\wotan} is based on its implementation in the {\tt scikit} package. It performs a spline-fit with knots spaced a distance $w$ apart, and uses the robust {\tt HuberRegressor} class to optimize the spline parameters. We found that {\wotan} implementation did not always converge, and made improvements by adding the {\tt StandardScaler} step to the {\tt scikit} pipeline and increasing the maximum number of iterations performed by the regressor ({\tt max\_iter=1000}). We also found that the default outlier threshold used by the regressor ({\tt epsilon=1.35}) is rather aggressive, down-weighing $35.8\%$ of the data points on average. We therefore opted to use a value of {\tt epsilon=3} instead, which down-weighs $0.3\%$ of the data points on average. We note that any such thresholds will likely need to be revisited once the real data is available, as the frequency of outliers may differ from the simulations analysed here.

\paragraph{Lowess.} The Lowess filter as implemented in {\wotan} de-trends the light curve by fitting a linear model to the data in a window of width $w$. Robustness is ensured by iteratively weighing the data, using a bisquare weight function for points whose residual are less than six times the median absolute deviation of the residuals and zero otherwise. In addition, points are weighted by the distance to the window centre using a tricube weight function. The use of the tricube weight function across the filter window means the optimal window size for the Lowess filters is larger than for the other tested filters. We fixed minor bugs related to the window edges and tricube weighting. 

\paragraph{YSD-Lowess.} The YSD-Lowess filter was introduced by \citet{Battley-2020} as a variant on the Lowess filter for the purpose of filtering the light curves of young, highly variable stars observed by \emph{TESS}. \citet{Battley-2020} started by detecting the locations of peaks and troughs in the light curve, then they remove data within an $0.2$~d window centred on each peak and trough from the light curve, before finally running a Lowess filter on the resulting light curve segments. In our implementation, the detection of the peaks and troughs is performed not on the raw light curve, but on the trend obtained by running a Lowess filter with a 6~h window. This step is necessary to suppress spurious detections of peaks and troughs in the high cadence PLATO data. We only consider peaks and troughs with a detected width between 7.5~h and twice the window-size ($w$) as wider features should be removed by normal Lowess de-trending. We mask data within an 0.4~d segment centred on the detected peaks and troughs and run Lowess de-trending on the resulting segments individually. Finally we fill the gaps in the resulting trend using cubic interpolation, avoiding the need to discard transit sized segments from the light curve.

\bigskip \noindent Figure \ref{fig:filters_quarter} shows the effect of the different filters on the 7th quarter of one of our simulated PLATO light curves, with a single transit occurring around $t \sim 602$~d. Even though this star is only moderately active, there is significant residual variability in the filtered light curves. The mean, biweight, and Lowess filters result in the highest residual amplitudes. The Huber spline and YSD-Lowess seem to perform better, but both still show significant residuals, and the YSD-filtered transit appears asymmetric (we discuss this in more depth in Sect.~\ref{sec:results}). Additional examples for a quieter and a more active star are shown in Appendix~\ref{app:supplementary-figures}. 

\begin{figure}[tbp!]
    \centering
    \includegraphics[width=1\columnwidth]{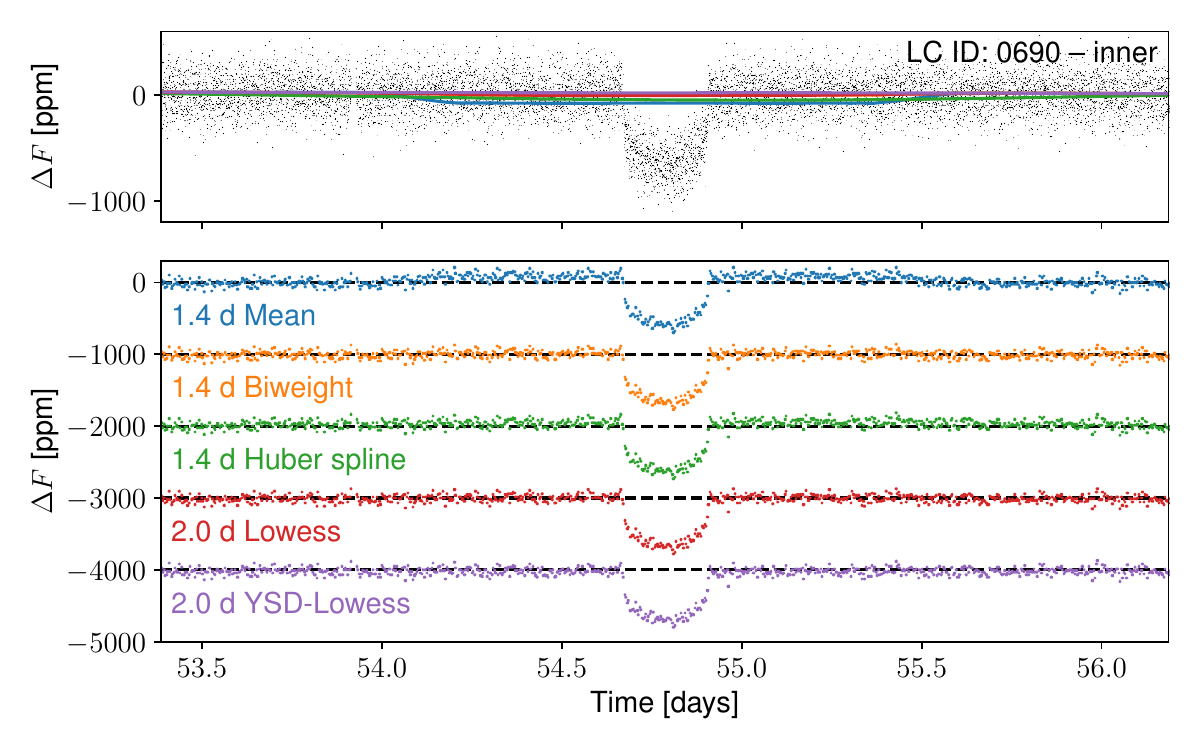}
    \caption{Same as Fig.~\ref{fig:filters_quarter}, but focusing on the effect the filters have on an individual transit.}
    \label{fig:filters_transit}
\end{figure}

Figure \ref{fig:filters_transit} shows the effect of the different filters on the shape of a single transit across a quiet star. The mean and Huber spline filter modify the signal shape, because the filter model absorbs some of the transit, which is then divided out. The other filters avoid this problem by using iterative outlier rejection to mask the transits. However, outlier rejection only prevents modification of the transit shape when the window is sufficiently wide ($w \gg T_{14}$) and when the individual transits have sufficient S/N.

\subsection{Transit search methods}
\label{sec:search}

We ran transit searches using three different algorithms: the published codes {\cetra} and {\nuance}, and a new code we have dubbed {\umbra}. Table~\ref{tab:search_defaults} provides an overview of the epoch, period and duration sampling employed by the different algorithms when producing the periodograms, and we discuss each in more detail in the following subsections. Independently of the results presented here an analysis of the same simulated light curves was performed using the AutoRegressive Planet Search methodology (ARPS, \citealt{Caceres-2019a, Melton-2024a}) which will be presented in Feigelson et al.\ (in prep.).

\begin{table*}[tbp!]
    \centering
    \caption{Default configurations for the transit search algorithms.}
    \label{tab:search_defaults}
    \begin{tabular}{llcclrll}
        \hline
        \hline
        Method & Filter & Cadence [s] & Epoch step [s] & Period grid & Period min & Duration grid & Duration limits \\
        \hline
        {\cetra} & various & 300 & 17.3 & Optimal\tablefootmark{a} & Roche limit\tablefootmark{a} & $0.02-1$~d & Yes\tablefootmark{b} \\
        {\umbra} & various & 300 & 60--300 & Optimal\tablefootmark{a} & $2w$ & {\umbra} ($OS=3$)\tablefootmark{b,c} & Yes\tablefootmark{b} \\
        {\nuance} & SHO GP & 600 & 600 & Optimal\tablefootmark{a} & 5~d & {\umbra} ($OS=1$)\tablefootmark{b} & No \\
        \hline
    \end{tabular}
    \tablefoot{ While we attempted to keep the parameter grids as similar as possible, allowances had to be made due to implementation differences and computational cost.
    \tablefoottext{a}{The period grids and the {\cetra} minimum period are all based on \citet{Ofir-2014}.}
    \tablefoottext{b}{The minimum and maximum overall durations in the grid and the period dependent duration limits were computed using the {\tls} package while assuming a $\pm20\%$ interval on the PIC stellar mass and radius. This choice presented issues as discussed in Sec.~\ref{sec:results-durations-limits}.}
    \tablefoottext{c}{{\umbra} computes multiple duration grids for subsets of the period range, allowing for a denser grid of durations while avoiding high computational cost.}
    }
\end{table*}

\subsubsection{{\cetra}}
\label{sec:search-cetra}

\citet{Hippke-2019-TLS} introduced the TLS algorithm and implemented it in the {\tls} package, which performs a least-squares search in pre-filtered light curve data using an approximate transit model. They demonstrated that this results in a significant improvement in sensitivity 
compared to the simple box shape employed by the BLS algorithm \citep{Kovacs-2002}, which was widely used previously. {\cetra}\footnote{\url{https://github.com/leigh2/cetra}} \citep{Smith-2025} is the default transit search algorithm within the Exoplanet Analysis System of the PLATO mission. It adopts many {\tls} features, but reduces the computational expense of the approach by separating it into a search for singular transit-like signals in linear time space, followed by a phase folding of the linear results to search for periodic signals. Furthermore, {\cetra} is written for NVIDIA GPUs using the CUDA framework, and benefits from their improved throughput for parallel operations compared to CPUs. Like {\tls}, {\cetra} requires input light curves to have been previously filtered. Residual stellar variability can lead to spurious detections, and this can hinder the detection of less significant real signals.

The linear search computes the likelihood ratio between a static flux model and a model containing a transit, with transit depth as an unrestrained free parameter, at every $t_0$ and duration grid-point. The periodic search then computes from these the joint-likelihood ratio of every $T_0$ (the $t_0$ of the reference transit), period and duration grid-point, assuming that all singular transit signals that comprise a periodic signal have a common transit depth. To reduce the number of relatively low-bandwidth data transfers between the GPU and the host machine, only the results of the highest likelihood ratio $T_0$, period and duration grid-point are returned for each period searched, which is sufficient to generate a periodogram. Periodic (or indeed singular) transit signals can be selected as either the maximum likelihood or maximum S/N. In this work we used the maximum S/N signal, which has the benefit that it naturally avoids selection of transits with a negative depth. Here we consider only the most significant signal in the light curve. During PLATO operations all signals greater than a pre-defined detection threshold will be reported, and this will tend to reduce the incidence of spurious signals masking real ones.

For the linear transit signal search, {\cetra} was provided with a transit model with impact parameter $b=0.32$, \mbox{$R_p=0.03R_*$}, and the input limb darkening parameters for each star. The transit duration grid is the {\cetra} default, 44 log-spaced durations from 0.02 to 1.0~d, with $t_0$ grid spacing of one per cent of the minimum duration (i.e. 17.3 seconds). {\cetra} permits extraction of mono-transits directly from the results of the linear search, but we did not do this to keep the approaches consistent between search methods. The periodic signal search used the {\tls} default period grid, which itself is based on the prescription of \citet{Ofir-2014}, given the input stellar mass and radius. When searching for a given period, only a subset of the duration grid is used, as per the {\tls} prescription, and for this we adopt $\pm{}20$~\% of the input stellar mass and radius as a range. Unless otherwise noted, we use {\cetra} v1.01.

\subsubsection{\umbra}
\label{sec:search-umbra}

{\umbra}\footnote{\url{https://github.com/talensgj/umbra}} is a newly developed periodogram-based transit search code, somewhat like BLS and TLS, but with two key differences: a filter-modified transit profile; and an updated $\Delta \chi^2$ detection statistic. {\umbra} will be described in detail in a future paper (Talens et al., in prep), but here we summarize these two important features. 

\begin{figure}[tbp!]
    \centering
    \includegraphics[width=1\linewidth]{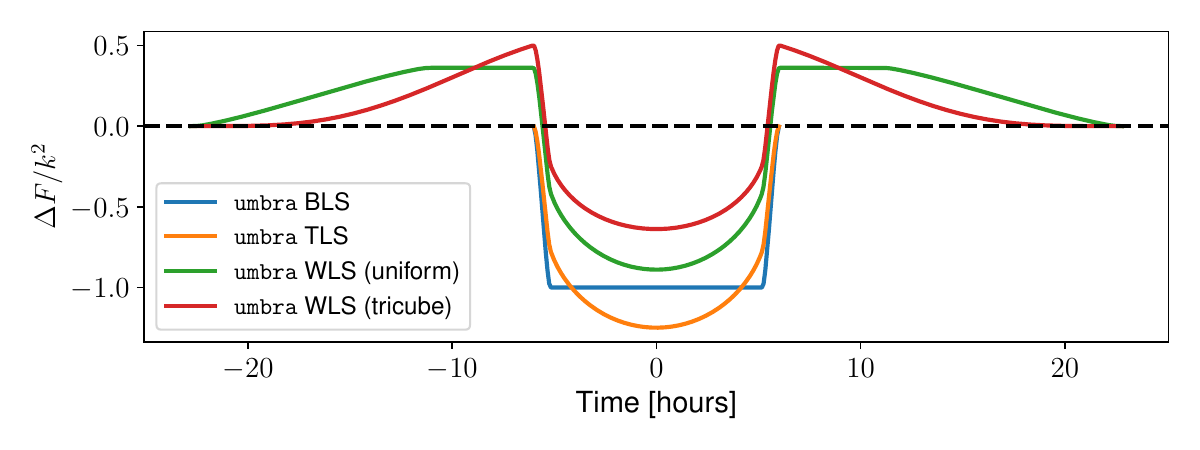}
    \caption{Different transit shapes that can be searched for with {\umbra}. BLS templates (blue) use a transit model with uniform limb-darkening, which mimics the template used by the BLS algorithm. TLS (orange) is a transit model with arbitrary limb-darkening, such as those used in {\tls} and {\cetra}. WLS (green and red) is a transit model with limb-darkening modified by a windowed mean filter with uniform (green) or tricube (red) weighting. All models had the out-of-transit flux level subtracted and were divided by $k^2=(R_p/R_\star)^2$.}
    \label{fig:warped_shapes}
\end{figure}

Light curve pre-filtering rarely leaves underlying transits untouched. Typically, some level of the transit signal is absorbed into the filter model and then divided out. Usually, this does not completely remove the signal, but it does modify, or ``warp'' it. This warping effect is visible in the mean and Huber spline filtered transits shown in Fig.\ref{fig:filters_transit}, but applies to other filters as well depending on the filter window size and transit S/N. {\umbra} can be run with transit templates that account for the warping effect that windowed filters have on the transit shape. These warped templates represents the next step in transit-template morphology beyond the box profile of BLS \citep{Kovacs-2002}, and the limb-darkened transit profile of TLS \citep{Hippke-2019-TLS}; we have dubbed this mode ``warped least-squares'' (WLS). {\umbra} implements versions of all three of these transit shape templates, as illustrated in Fig.~\ref{fig:warped_shapes}. First, a transit with uniform limb-darkening is shown as a BLS template, but note that this differs from the standard box template in that we included ingress and egress time. Next, we show the TLS template, which includes realistic limb-darkening. Finally, the TLS template is passed through a 1.4~d mean filter and a 1.4~d tricube weighted mean filter to give two examples of WLS transit templates. All of the transits shapes have had the out-of-transit flux level removed and have been divided by $k^2=(R_p/R_s)^2$. The warping effect of the filtering on the transit models is evident -- the peak depth of the transit is reduced, and flux in the wings and outside the transit is increased. The exact shape of the warping is filter-method dependant. Currently {\umbra} only implements a mean filter with different weighting schemes to compute the WLS transit templates, as robust estimators cannot be trivially applied to a noise-free model. Even with that limitation {\umbra} with WLS templates outperforms {\cetra} on all tested filters, as we will see in Sect.~\ref{sec:results-single-filters}. Generating the warped transit shapes is expensive at short periods, where the filter window $w$ may contain parts of more than one transit event. Only periods $P > 2w$ were searched to avoid this regime. We do not expect this to have a major impact on the results as these signals can still be detected at multiples of the true period, and indeed eight of 12 inner signals with $P \leq 2.8$~d are recovered in the 1.4~d Huber spline filtered light curves. Unless otherwise noted, we use WLS transit templates generated with uniform weights for the biweight and Huber spline filters, and WLS templates generated with tricube weights for the Lowess and YSD-Lowess filters.

When BLS was first introduced \citep{Kovacs-2002}, it used $\Delta \chi^2$, the difference between the $\chi^2$ for the transit model under consideration, and that for flat line $\chi^2_0$, as the main detection statistic. The well established Lomb-Scargle periodogram for sinusoidal oscillations uses a similar statistic. In its simplest form, the BLS periodogram power is defined as
\begin{align}
\label{eq:bls_stat}
   \mathcal{P}(P) = \max_{T_0, T_{14}} \left( \frac{\chi^2_0 - \chi^2(P, T_0, T_{14})}{\chi^2_0} \right),
\end{align}
where the constant denominator is usually omitted. Often the $\max$ operation is limited to the $T_0$, $T_{14}$ producing best-fit transits with physical depths ($\delta < 0$). The resulting power is usually empirically post-processed into the signal residue, which has many variants \citep{Kovacs-2002, Ofir-2014, Hippke-2019-TLS}. The {\umbra} algorithm instead uses
\begin{align}
\label{eq:umbra_stat}
   \mathcal{P}(P) =& \max_{T_0, T_{14}} \left(\frac{\min_{T_0,\delta > 0}[\chi^2(P, T_0, T_{14})] - \min_{\delta < 0}[\chi^2(P, T_0, T_{14})]}{\min_{T_0,\delta > 0}[\chi^2(P, T_0, T_{14})]} \right )\\
   = & \max_{T_0, T_{14}} \left(\frac{\chi^2_{+}(P,T_{14}) - \chi^2_{-}(P, T_0, T_{14})}{\chi^2_+(P,T_{14})}\right)
\end{align}
where $\chi^2_-$ is the best-fit physical transit model ($\delta < 0$), and $\chi^2_+$ the best-fit unphysical model ($\delta \geq 0$) at the same $P$, $T_{14}$ as $\chi^2_-$. The idea being that the $\delta{}<0$ solutions contain both noise and genuine transit signals, while the $\delta{}\ge{}0$ solutions contain only noise at the same period and duration, providing a data-driven noise reference. Unlike $\chi^2$ and $\chi^2_0$ in Equation~\ref{eq:bls_stat}, the models used to compute $\chi^2_+$ and $\chi^2_-$ have the same number of free parameters. Comparing the best-fit physical model to this noise reference at the same period and duration results in a conservative, self-normalising statistic that suppresses spurious detections caused by noise and systematics, and removes the need for any empirical post-processing. An example periodogram is included in Appendix~\ref{app:supplementary-figures}.

When running {\umbra} we use an epoch grid with steps tuned to align the shortest transit shapes to within a fraction of the ingress time, but limiting the epoch step to a range of 1--5~mins depending on the orbital period to avoid over/undersampling of the transit shape. To build the duration grid, {\umbra} uses functions from the {\tls} package to compute the longest and shortest durations as a function of period assuming a $\pm20\%$ range centred on the PIC stellar mass and radius. These values are then used to group periods so that within the group $T_{14,\max}/{P_{\min}} < 0.15$ and duration limits are recomputed for the period group. A duration grid is then built for each period group by iteratively adding $(T_{14} - T_{23})/OS$ to the current duration to get the next duration, starting from the shortest duration. The {\umbra} searches presented here use $OS=3$. Finally WLS transit templates are built for the duration grid of each period group while accounting for both the exposure time and exposure cadence of the observations. The transit shapes are nominally evaluated with $b = 0$ and $k^2 = 5000$~ppm. In this work versions 0.2.0 through 0.5.0 of {\umbra} were used.

\subsubsection{\nuance}
\label{sec:search-nuance}

{\nuance}\footnote{\url{https://github.com/lgrcia/nuance}} \citep{Garcia-2024} is a new transit search algorithm that uses GPs combined with a low-order polynomial to account for stellar variability. Unlike {\cetra} and {\umbra} this GP model is evaluated simultaneously with the analytic transit model of \citet{Protopapas-2005}. The simultaneous evaluation of the transit model and the GP should make {\nuance} superior at both accounting for stellar variability and disentangling it from transit signals, at the cost of being significantly more computationally expensive. For this work we made several implementation choices to minimize the computational cost, as outlined below.

A typical {\nuance} implementation starts by optimizing the GP hyperparameters, followed by a linear search across a grid of epochs and durations, before finally performing a periodic search using the linear search results. In our implementation we start by optimizing a simple harmonic oscillator (SHO) GP kernel on the full two year PLATO light curve binned to 10-minute cadence. The GP has the following hyperparameters: $\sigma^2_{\rm{SHO}}$ the variance of the GP kernel, $\omega = 2\pi/P_{\rm{SHO}}$ the angular frequency of the oscillator, $Q$ the quality factor of the oscillator, and $\sigma_{\rm{jit}}$ a jitter term added to the photometric uncertainties on the diagonal. For the GP mean model we use a $4^{\rm th}$ order polynomial on each quarter of observations, expressed as a block diagonal matrix. When optimizing the hyperparameters on versions of the light curves with and without transit signals, we find that the presence of transits pushes the ratio of the hyper parameters $\omega/Q$ to larger values so we impose a constraint $\ln \omega/Q < 0.5$. We also constrain $0.1~\rm{d} < P_{\rm{SHO}} < 2S$, where $S$ is the two year baseline of the observations. Once the hyper parameters have been optimized, they are fixed for the remainder of the analysis. 

We perform the linear search on each quarter of observations individually to save computational time. The epoch grid is set to have a 10~min resolution, and the shortest and longest duration are computed using functions from the {\tls} package, using a $\pm20\%$ range on the PIC stellar parameters. The duration grid is filled using the {\umbra} method of using $(T_{14} - T_{23})/OS$ increments, using $OS=1$ to limit the number of durations and save computational time. Finally, the linear searches are combined and the periodic transit search is performed; no period-dependent duration limits were imposed at this stage. Even with the higher binned cadence and reduced duration grid, the period search still proved prohibitively expensive on the long, densely sampled PLATO light curves, so we restricted the search to periods $P > 5$~d to save further computation time. As with {\umbra}'s WLS we do not expect this to have a major impact on the results, and indeed 19 of 34 simulated inner signals with $P \leq 5$~d are detected at twice the injected period. 

\section{Results}
\label{sec:results}

The full parameter space of transit samples, filters, filter windows, and transit search codes is too large to be fully explored without incurring excessive computational cost. Instead we adopted a hierarchical approach to exploring the parameters space. For {\cetra} and {\umbra} an initial search was performed on the more challenging outer simulations only, and using only four window sizes for each of the four filters. In this initial search the most signals were recovered with {\umbra} on the 2.0~d Lowess filter. However, the Huber spline filter more consistently outperformed other filters across a range of window sizes with both {\cetra} and {\umbra}. Based on this we ran new Huber spline searches on the inner and outer simulations using eight window sizes. {\nuance} was run on both the inner and outer simulations.

To establish how many signals can be detected in the absence of long-term variability and systematics we also ran each code on the unfiltered simple light curves of both the inner and outer simulations. {\cetra} was run using the same parameters as the filtered lightcurves; {\umbra} used the TLS templates as no filter had warped the transit shapes; and {\nuance} used a diagonal term for the kernel and a 0th order polynomial for the mean model in each quarter to avoid over-fitting by the full GP model. 

Signals are considered correctly retrieved if the recovered period satisfies:
\begin{align}
\label{eq:crit1}
    (P_{\rm{inj}} - n P_{\rm{rec}})N_{tr} < & \frac{T_{14,\rm{inj}}}{2}
\end{align}
for $n=1/2,1,2,3$, i.e.\ if the error on the transit time accumulated over the full light curve due to the error on the period is less than half of the transit duration.

While none of the transit search algorithms considered here were run in a dedicated mono-transit mode, alignment of individual transit events with the larger quarter gaps in the observations can still produce peaks in the periodogram. These peaks are especially relevant for light curves with few transit events. As such for light curves with $N_{tr}=1,2$, we also consider a transit recovered if:
\begin{align}
\label{eq:crit2}
    |T_{0,\rm{inj}} + mP_{\rm{inj}} - T_{0,\rm{rec}}| < & \frac{T_{14,\rm{inj}}}{2}
\end{align}
with $m=0$ for mono-transits and $m=0,1$ for duo-transits. This second criterion only applies to the outer sample, since the inner sample contains no mono- or duo-transits. The number of signals recovered in all the searches described above are listed in Table~\ref{tab:overview}.

We do not apply any thresholds to the strength of the periodogram peaks, since all three transit search algorithms provide different statistics, and the analysis needed to obtain fair thresholds for each is beyond the scope of this work, but see App.~\ref{app:thresholds}.

\subsection{Filter comparison}
\label{sec:results-single-filters}

\begin{figure*}[tbp!]
    \centering
    \includegraphics[width=1\columnwidth]{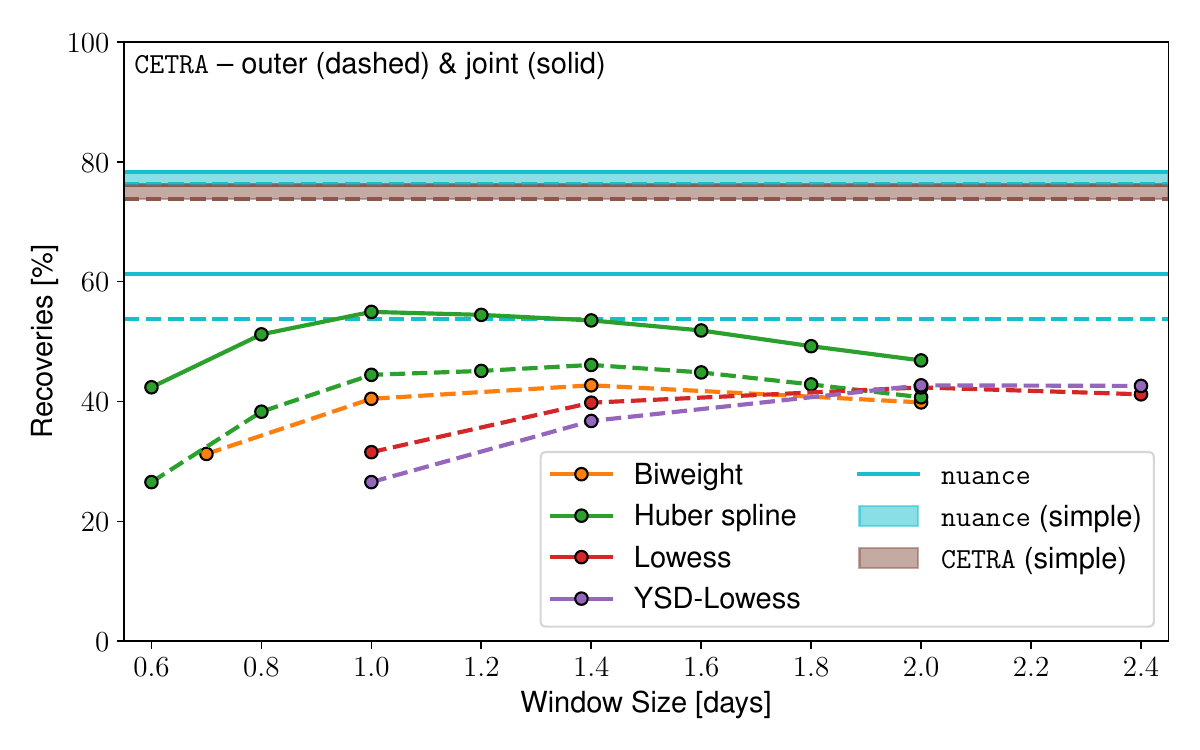}
    \includegraphics[width=1\columnwidth]{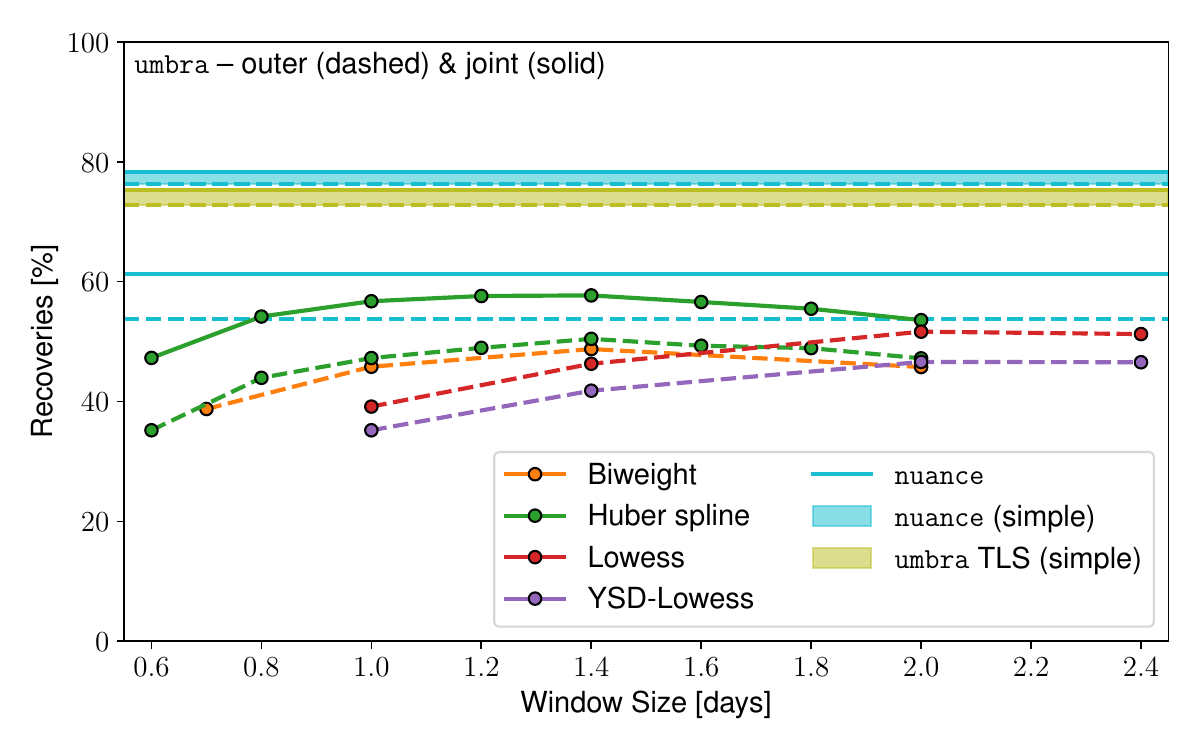}
    \caption{Percentage of transits recovered by the different algorithms and filters as a function of filter window size. Results are shown for the outer sample (dashed) and the joint sample (solid). Left: signals recovered by {\cetra} in the different filters (line scatter). Right: same as left but for {\umbra}. In both panels the horizontal lines show the {\nuance} (cyan) recovery percentage which is independent of window size. Coloured bands indicate the range of recovery percentages obtained on the simple lightcurves by {\cetra} (brown), {\umbra} (yellow), and {\nuance} (cyan).}
    \label{fig:search_results}
\end{figure*}

The percentage of signals recovered in the different transit searches performed are shown in Fig.~\ref{fig:search_results}. The three transit search codes recover a similar percentage of transits in the simple joint simulations, 78.3\% with {\nuance}, 76.2\% with {\cetra} and, 75.3\% with {\umbra} TLS. 

On the full simulated lightcurves {\nuance} performs best, recovering 53.8\% of signals in the outer transit sample, and 61.2\% in the joint sample. It is followed by {\umbra} which recovers 51.7\% of signals in the outer sample with the 2.0~d Lowess filter, and 57.7\% of signals in the joint sample with the 1.4~d Huber spline. Finally, {\cetra} recovers 46.1\% of outer signals with the 1.4~d Huber spline, and 55.0\% of joint signals with the 1.0~d Huber spline.

While {\nuance} performs best, the improvement in the number of recoveries is small when weighed against the significant computational cost of evaluating the GP kernel. With the extra freedom afforded by the GP kernel we might have expected {\nuance} to close more of the gap between the results obtained on the full light curves, and what is theoretically possible in the absence of long-term variability as shown by the simple light curves. It is possible that {\nuance} can be improved by using a different GP, or by performing the linear search on the full two year light curve, but this is beyond the scope of this work.

The improvement of {\umbra} over {\cetra} is small, and {\umbra} also comes at additional computational cost. However, if the WLS templates were to be implemented in {\cetra} this cost would be minimised. 

With {\umbra} there is a more consistent difference between the Lowess and YSD-Lowess filters than with {\cetra}, especially at longer windows where these filters perform best. This improved performance of the Lowess filter compared to the YSD-Lowess filter is likely because the interpolation across the peaks and troughs used in the YSD-Lowess filter does not always have the same effect on transit features. 

\begin{figure*}[tbp!]
    \centering
    \includegraphics[width=1\linewidth]{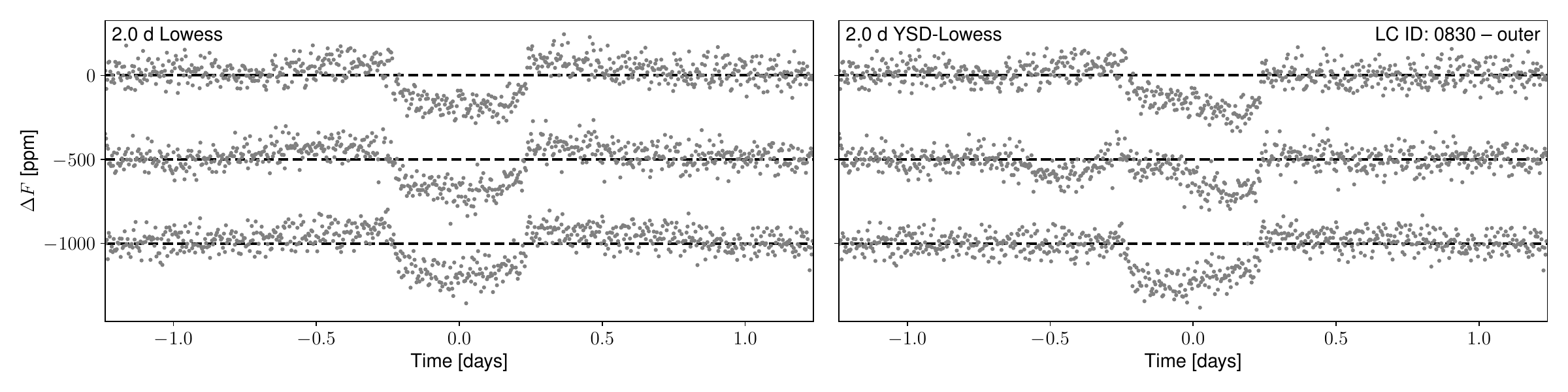}
    \caption{Transits in a Lowess filtered light curve (left) compared to the same light curve filtered with YSD-Lowess (right). Lowess has modified the transit shapes and surrounding baseline, but all three transits have similar filtered shapes. By contrast YSD-Lowess has less of an effect on the baseline, but has severely distorted the middle transit.}
    \label{fig:ysd_lowess_example}
\end{figure*}

Fig.~\ref{fig:ysd_lowess_example} shows the effect of the Lowess and YSD-Lowess filters on three transits in the same lightcurve. Both filters distort the shape of the transit, but for Lowess this distortion is consistent across the three transits and {\umbra}'s WLS can account for it. YSD-Lowess also distorts the transit shapes, but the distortion differs in each transit. The 2nd transit in particular is especially affected. This inconsistent behaviour is caused by the interpolation across cut segments, which {\umbra}'s WLS cannot account for. As a result, the Lowess filter performed better than the YSD-Lowess filter in this case. 

\cite{Canocchi-2023} found somewhat better performance with the YSD-Lowess filter than we have done here relative to the other filters they tested. This difference is likely because they focused on larger planets on shorter orbits around active stars. Implementation differences in the YSD-Lowess filter between their work and ours may also be contributing.

\subsection{Combining filter algorithms}
\label{sec:results-combine-filters}

\begin{figure*}[tbp!]
    \centering
    \includegraphics[width=1\columnwidth]{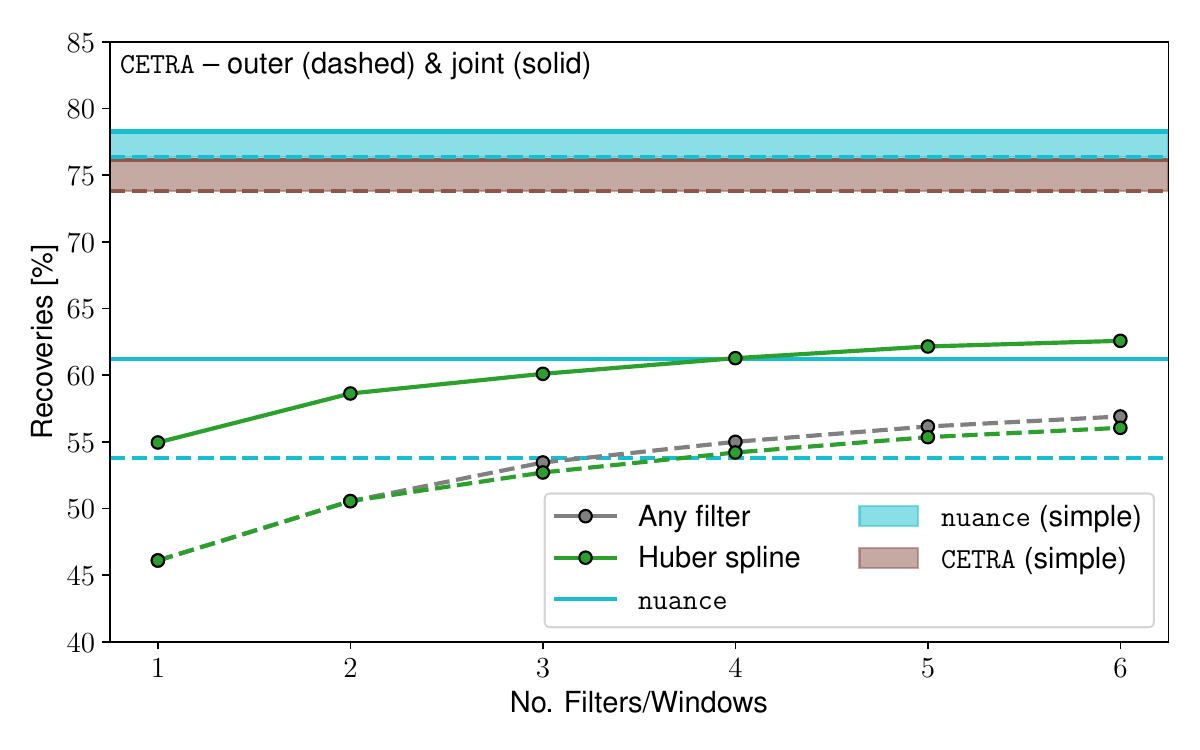}
    \includegraphics[width=1\columnwidth]{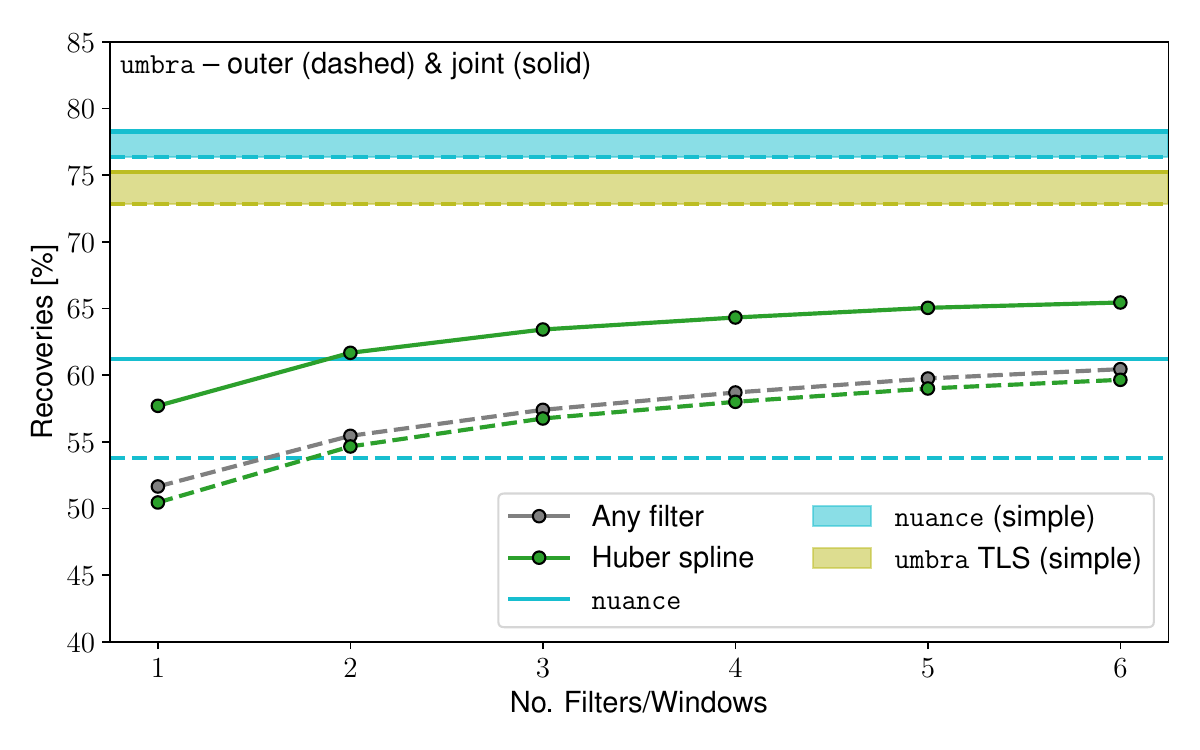}
    \caption{Percentage of transits recovered by combining different light curve filters and different window sizes. Results are shown for the outer sample (dashed) and the joint sample (solid). Left: signals recovered by {\cetra} in the optimal union of $N$ different filter-window combinations. Right: same as left but for {\umbra}. In both panels the horizontal lines show the {\nuance} (cyan) recovery percentage, and coloured bands indicate the range of recoveries obtained on the simple lightcurves by {\cetra} (brown), {\umbra} (yellow), and {\nuance} (cyan).}
    \label{fig:filter_combos}
\end{figure*}

In principle, the PLATO pipeline could run multiple combinations of light curve filters and transit search algorithms, subject to computational constraints. In this section we explore how many unique signals {\cetra} and {\umbra} could recover in the union of multiple filters and windows. We test all permutations of one to six filters and record the optimal combination of filters and number of recovered signals. We perform this test using all filter-window combinations in the outer sample, and only Huber spline filters in the inner, outer and joint samples. The results are shown in Fig.~\ref{fig:filter_combos} and listed in Table~\ref{tab:filter-combos}.

With six filter-window combinations we gain approximately 4-5~percentage points (\%pt.) in the inner sample, 9-11~\%pt. in the outer sample,  and 7.6~\%pt. in the joint sample. {\cetra} recovers more signals than {\nuance} with four filter-window combinations, and {\umbra} exceeds {\nuance} with two filter-window combinations. In both cases this represents less computational cost than running {\nuance}. When comparing the performance on the outer sample, there is only a minimal difference between using any of the tested filters and windows or just the Huber spline with different windows. At most a single non-Huber spline filter is ever included when allowing all tested filters. When considering 1--2 filter-window combinations {\umbra} includes the 2.0~d Lowess filter, for 3--6 filter-window combinations {\umbra} and {\cetra} include the 2.0~d YSD-Lowess filter. The advantage gained by including filters other than the Huber spline is less than 1~\%pt., and because of the inconsistent effect of the YSD-Lowess filter we do not recommend including a filter other than the Huber spline. 

\subsection{Search comparison}
\label{sec:results-comparison}

With a single filter {\nuance} performs best, recovering 61.2\% of signals in the joint sample. It is followed by {\umbra} with 57.7\% in the 1.4~d Huber spline and {\cetra} with 55\% of signals on the 1.0~d Huber spline. The union of al three transit search algorithms allows the recovery of approximately 65\% of transit signals. A significant increase compared to using any of the individual algorithms.

To determine in which part of parameter space each search algorithm perform best, and which algorithms would complement each other, we compare the signals recovered in the joint sample as a function of transit and stellar parameters, respectively. {\cetra} is used as the point of reference, and we use the 1.4~d Huber spline filter for {\cetra} and {\umbra} so we can focus on differences between the search algorithms.

\begin{figure*}[tbp!]
    \centering
    
    \includegraphics[width=1\linewidth]{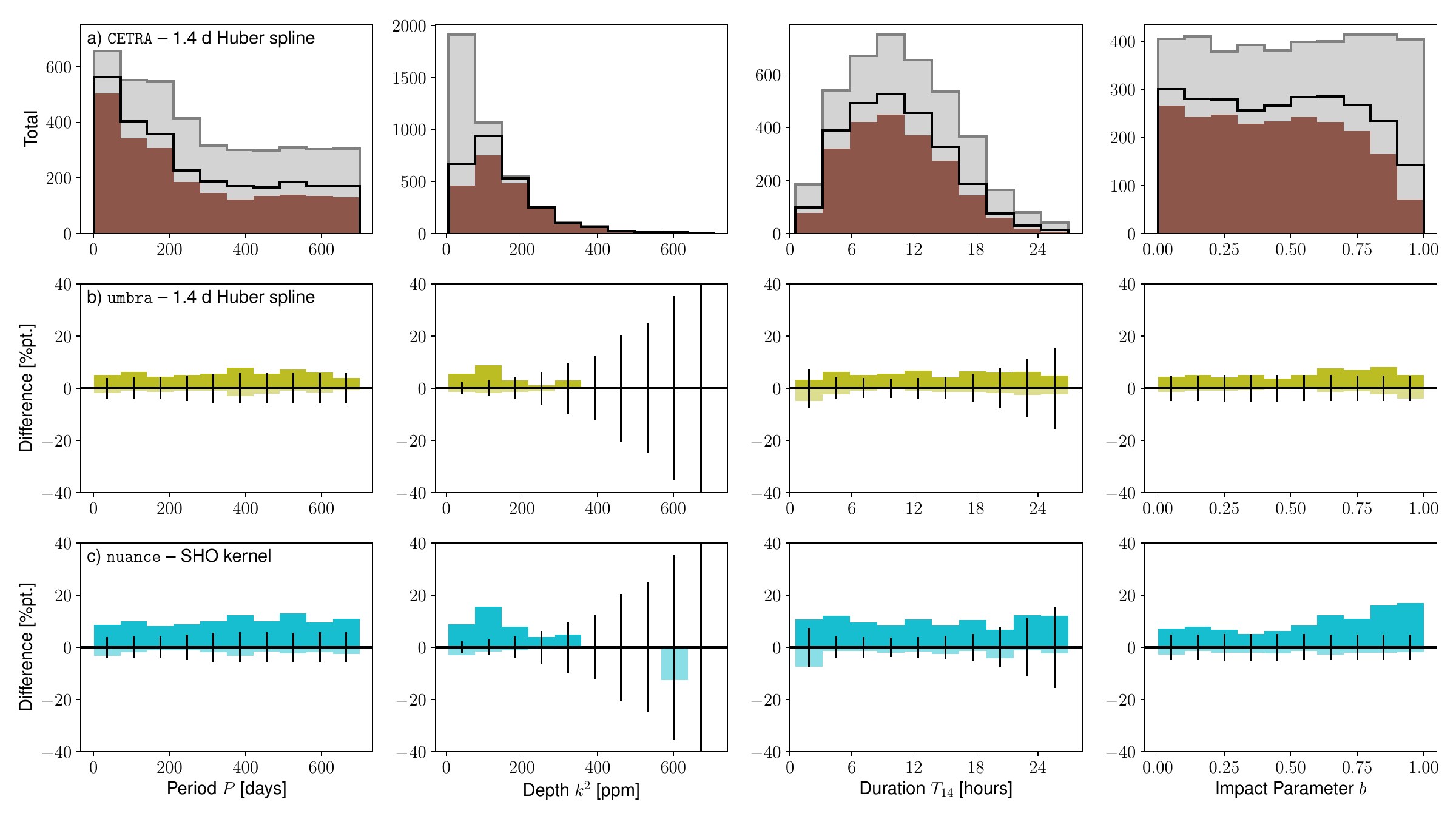}
    \caption{Transit recovery rates as a function of the transit parameters in the joint sample for the {\cetra}, {\umbra}, and {\nuance} algorithms. {\cetra} and {\umbra} are shown for the 1.4~d Huber spline filter, to allow a like-for-like comparison. Row a) shows the total number of simulated signals (grey), the signals recovered by {\cetra} (brown), and the union of signals recovered from all three algorithms (black). Rows b) and c) show the difference between {\umbra} and {\cetra} (yellow) and between {\nuance} and {\cetra} (cyan), respectively. The differences are shown in percentage points (\%pt.), with counting errors indicated by the black vertical lines. Signals recovered by {\umbra}/{\nuance} but missed by {\cetra} are indicated by positive values (solid), and the reverse by negative values (faded).}
    \label{fig:search_comparison_transit}

    \includegraphics[width=1\linewidth]{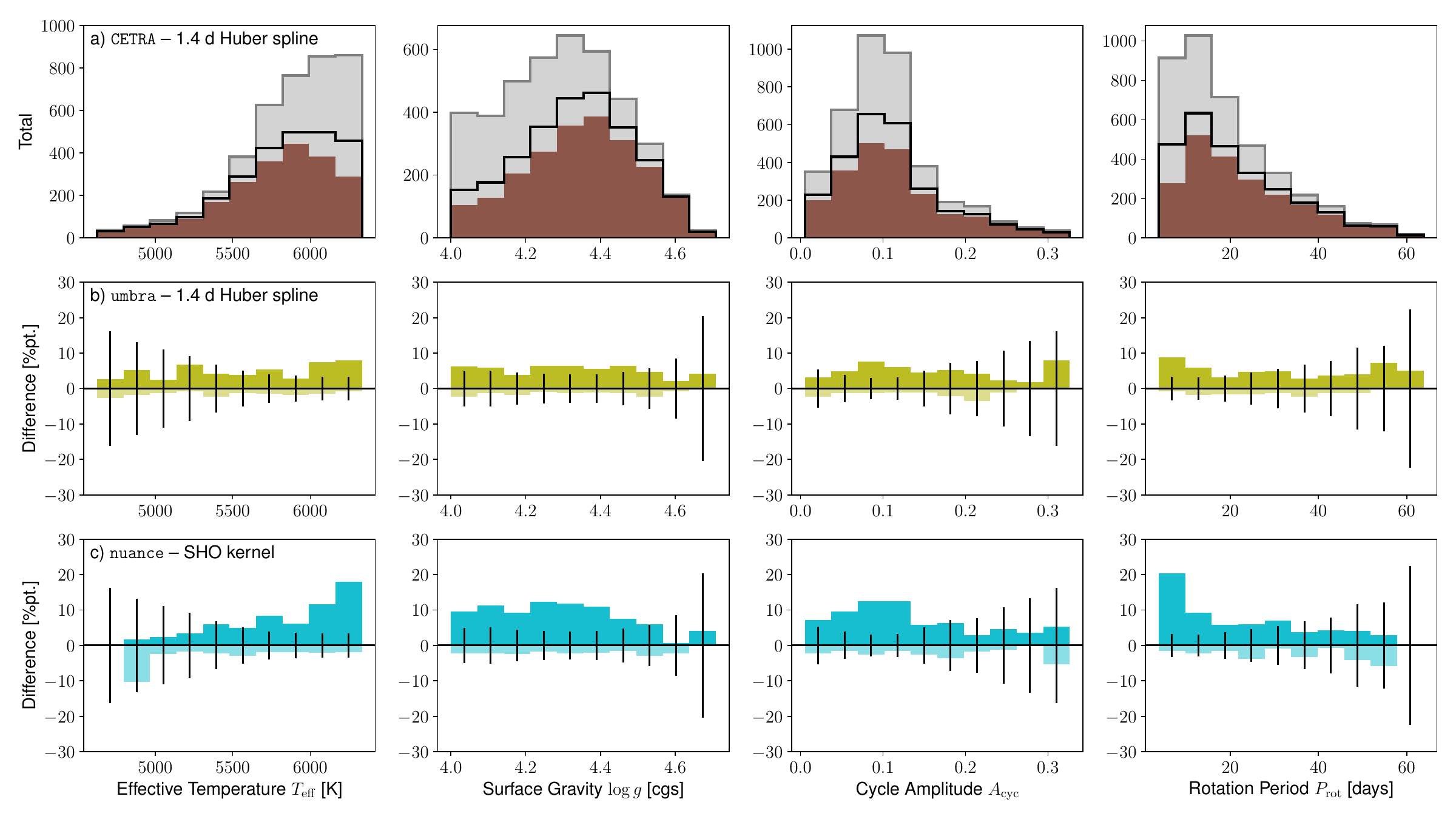}
    \caption{Same as Figure~\protect\ref{fig:search_comparison_transit}, but as a function of stellar parameters.}
    \label{fig:search_comparison_stellar}
    
\end{figure*}

Fig.~\ref{fig:search_comparison_transit} shows that {\umbra} and {\nuance} perform better than {\cetra} across transit parameter space. Both algorithms show improved recoveries of shallow transits, and {\nuance} also performs better at high impact parameters. Improvements as a function of period and duration are likely driven by the other parameters. The number of signals detected by {\cetra} but missed by {\umbra} and {\nuance} are smaller than the error bars, only the shallow transits missed by {\nuance} are marginally significant. Fig.~\ref{fig:search_comparison_stellar} reveals that improved recoveries in {\nuance} are driven by hot, rapidly rotating stars. The same effect is present to a lesser degree for the {\umbra} results. There is significant correlation between $T_{\rm{eff}}$ and $P_{\rm{rot}}$ (as show in Fig.~\ref{fig:correlations}).

\subsection{Further tests}
\label{sec:results-further-tests}

\subsubsection{The effects of binned cadence}
\label{sec:results-bin-filter}

\begin{table}[tbp!]
    \centering
    \caption{Filtering and binning order and binned cadence test results.}
    \label{tab:overview_filter_bin}
    \begin{tabular}{lcclr}
        \hline
        \hline
         Scheme & Bin Size & Filter Time & Search & outer \\
         & [s] & [s] & & 2000 \\
         \hline
         Filter $\rightarrow$ Bin & 150 & 103.6 & {\umbra} & 1008 \\
         Filter $\rightarrow$ Bin & 300 & 142.9 & {\umbra} & 1009\tablefootmark{a} \\
         Filter $\rightarrow$ Bin & 600 & 103.1 & {\umbra} & 1009 \\
         Bin $\rightarrow$ Filter & 150 & 21.7 & {\umbra} & 1006 \\
         Bin $\rightarrow$ Filter & 300 & 7.6 & {\umbra} & 1008 \\
         Bin $\rightarrow$ Filter & 600 & 4.1 & {\umbra} & 1011 \\
         \hline
         Filter $\rightarrow$ Bin & 150 & ... & {\cetra} & 922 \\
         Filter $\rightarrow$ Bin & 300 & ... & {\cetra} & 922\tablefootmark{a} \\
         Filter $\rightarrow$ Bin & 600 & ... & {\cetra} & 921 \\
         Bin $\rightarrow$ Filter & 150 & ... & {\cetra} & 921 \\
         Bin $\rightarrow$ Filter & 300 & ... & {\cetra} & 920 \\
         Bin $\rightarrow$ Filter & 600 & ... & {\cetra} & 920 \\
         \hline
    \end{tabular}
    \tablefoot{All tests were performed using the 1.4~d Huber spline filter on the outer sample. \tablefoottext{a}{The default of filtering the 25~s cadence light curve before binning to 300~s used throughout this work.}}
\end{table}

So far, all filters were run on the 25~s cadence light curves, before binning to 300~s cadence for the transit search. We now explore binning the 1.4~d Huber spline filtered light curves to 150 and 600~s. Furthermore, we explore binning the light curves to 150, 300, and 600~s cadence before running the 1.4~d Huber spline filter. The resulting light curves were searched using {\cetra} and {\umbra}, and the results are shown in Table~\ref{tab:overview_filter_bin}.

We find that {\cetra} and {\umbra} recover approximately the same number of signals, regardless of the binned cadence and whether the light curves were filtered before or after binning. There is also high overlap between the recovered signals, {\cetra} recovers the same 913 signals in all cases and {\umbra} the same 998 signals.

The differences are too marginal to choose a preferred bin-size among the ones tested here, but do suggest that we can reduce the cadence before filtering. Doing so would also significantly reduce the computational cost of running the light curve filters (Sec.~\ref{sec:results-performance}) from 142.9~s on the 25~s cadence light curve, to 7.6~s on the 300~s cadence light curves. No bin-size or GP choice tests were performed with {\nuance} due to the computational cost of the algorithm. PLATO light curves should not be binned to higher cadences than the ones tested here without careful treatment of the $\sim20$~min data gaps resulting from spacecraft operations.

\subsubsection{Transit duration limits}
\label{sec:results-durations-limits}

The extremes of the duration grid searched varied between the different transit search algorithms (see Table~\ref{tab:search_defaults}). {\cetra} used fixed extremes for the duration grid while {\umbra} and {\nuance} used the functions provided in the {\tls} package to compute extremes based on the host star parameters. {\cetra} and {\umbra} then apply period dependent limits based on the {\tls} package, while {\nuance} does not use period dependent limits. Duration limits were computed for each star with the {\tls} package using a $\pm20\%$ interval around the PIC stellar mass and radius. 

During the preparation of this manuscript we discovered that while the default stellar parameter intervals used by {\tls} account for a wide range of possible transit durations, small parameter intervals cannot be used as physical priors for a specific star for two reasons. First, when {\tls} converts the mass and radius intervals to duration limits the mass limits are swapped. Second, {\tls} does not account for the effects of the impact parameter or the eccentricity on the transit duration. 

It is possible to analytically derive the extreme values that the transit duration can take for a specific host star, as demonstrated in \citet{Talens-2025}. Implementing these equations for the extremes of the duration grid, as well as the period dependent limits, and re-running all affected {\cetra}, {\umbra}, and {\nuance} runs would have been too computationally expensive. Instead we present here a simple test performed with {\cetra} v1.03, which implements the \citet{Talens-2025} duration limits.

In this test we ran {\cetra} on the 1.0 and 1.4~d Huber spline filtered light curves. {\cetra} used the same duration grid as before, running from $0.02-1$~d, but computed duration limits for each host star based on \citet{Talens-2025} while assuming circular orbits and using a $\pm20\%$ interval on the mass and radius from the PIC. In these runs we recovered 47.4(68.5)\% signals in the outer(inner) sample with the 1.0~d Huber spline filter, and 48.5(63.8)\% of signals with the 1.4~d Huber spline filter. For comparison, we recovered 44.5(65.5)\% and 46.1(61.0)\% of signals with the {\tls} limits in the 1.0 and 1.4~d filter respectively. In other words, the extended duration grid yields an improvement of several percentage points in both filters.

Further tests using the duration limits presented in \citet{Talens-2025} will be needed, especially to investigate the impact of including short duration at long periods on the periodogram, but are beyond the scope of this work. 

\subsubsection{Recovered parameter biases}
\label{sec:results-parameter-biases}

\begin{figure}[tbp!]
    \centering
    \includegraphics[width=1\linewidth]{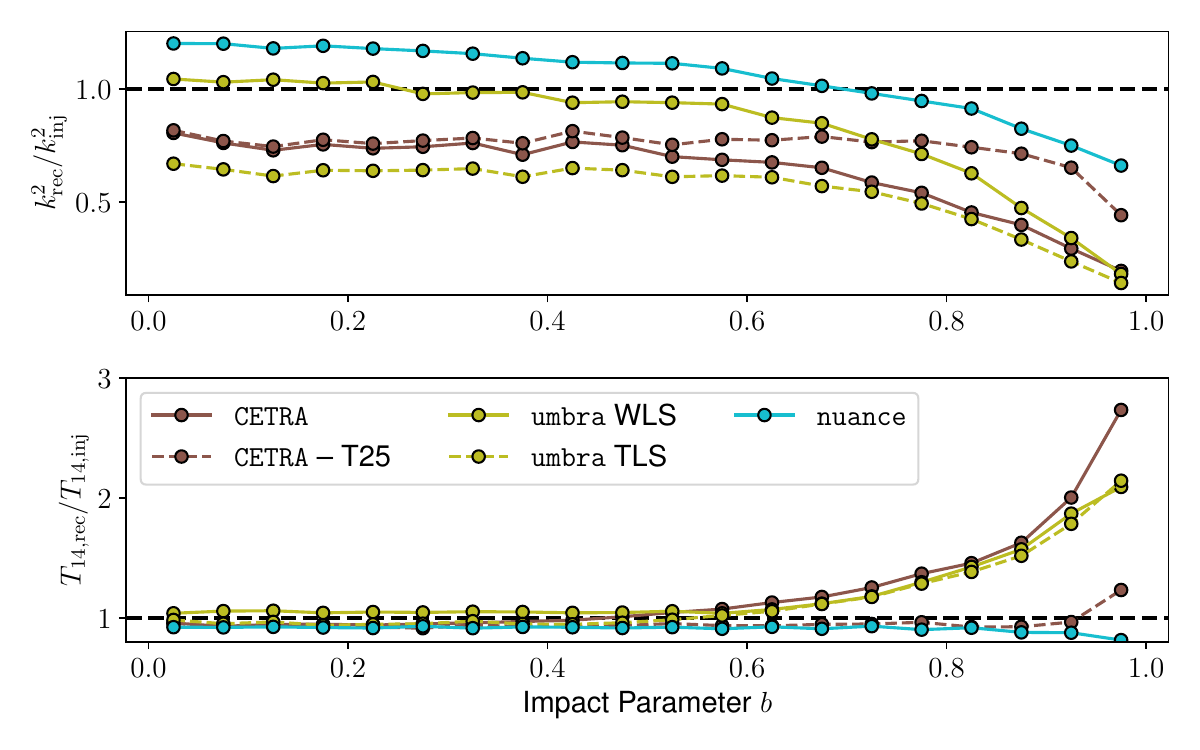}
    \caption{Biases in the recovered transit depth and transit durations for different runs of the transit search algorithms. Shown are the differences between {\cetra} with the default configuration (yellow); {\cetra} with the \citet{Talens-2025} duration limits (dashed yellow); {\umbra} with TLS templates, {\umbra} with WLS templates; and {\nuance}. Top: ratio of the recovered and injected transit depths. Bottom: ratio of the recovered and injected transit durations. Unbiased recoveries are indicated by the dashed black line.}
    \label{fig:biases}
\end{figure}

While the transit parameters recovered by the different transit search algorithms do not need to be particularly precise, those with lower bias are doing a better job of fitting the transits and should therefore be capable of detecting lower S/N signals successfully. Figure~\ref{fig:biases} shows the biases in the depths and durations of the successfully recovered transits for the tested algorithms. We show {\umbra} using both WLS and TLS templates, and {\cetra} using the {\tls} duration limits and the \citet{Talens-2025} duration limits. 

We find that both the {\umbra} runs and the original {\cetra} run are biased towards long duration when recovering high impact parameter transits. This bias is the direct result of the duration limits employed, which do not include sufficiently short durations to represent these high impact transits. The {\cetra} - T25 results do not have this bias as more appropriate limits were used. \nuance\ did not use any limits, thereby avoiding duration biases at the potential cost of excess noise in the periodogram.

Both {\cetra} runs and the {\umbra} run using TLS templates underestimate the planet radius, due to the filters reducing the transit depth compared to the baseline. {\umbra} using WLS templates avoids this bias, correctly recovering the planet radii. {\nuance} overestimates the planet radius, possibly as a result of using the \citet{Protopapas-2005} transit model. All algorithms underestimate the planet radius at high impact parameters, which is expected as the templates do not account for the effect of the impact parameter on the transit shape. Overall it looks like a combination of WLS templates and \citet{Talens-2025} transit duration limits would produce the least bias, and thus the highest sensitivity.

\subsection{Computational performance}
\label{sec:results-performance}

\begin{figure*}[tbp!]
    \centering
    \includegraphics[width=1\columnwidth]{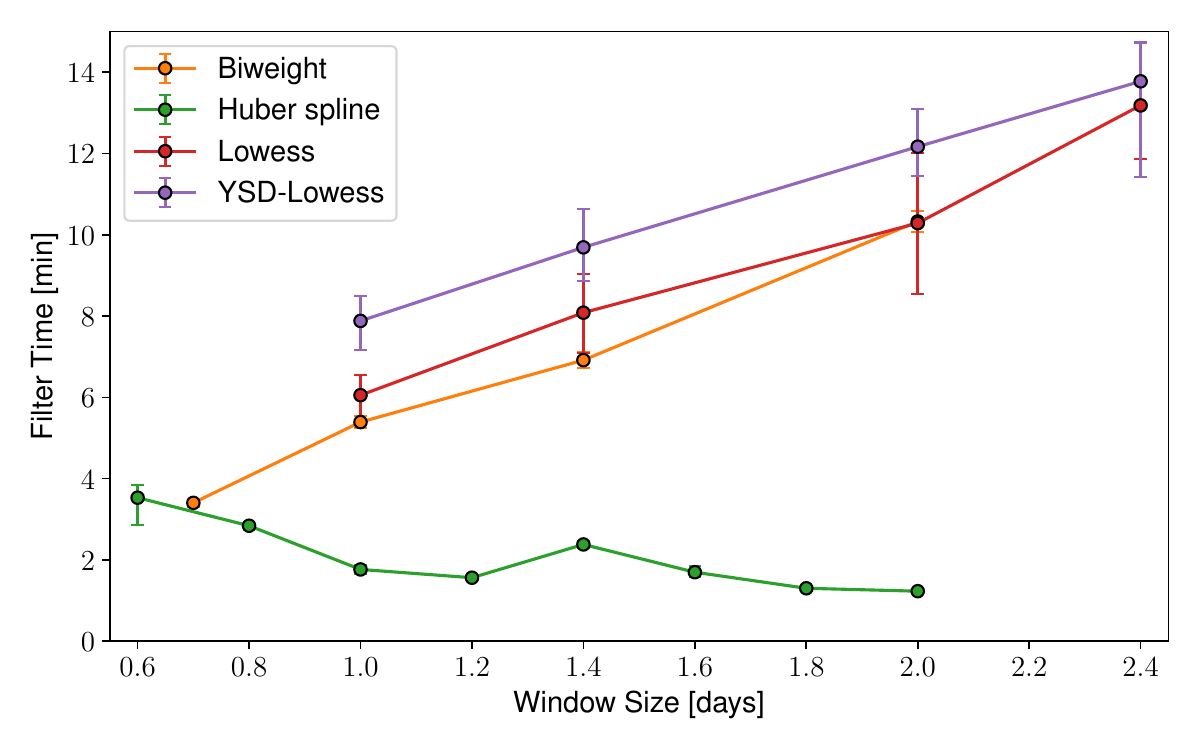}
    \includegraphics[width=1\columnwidth]{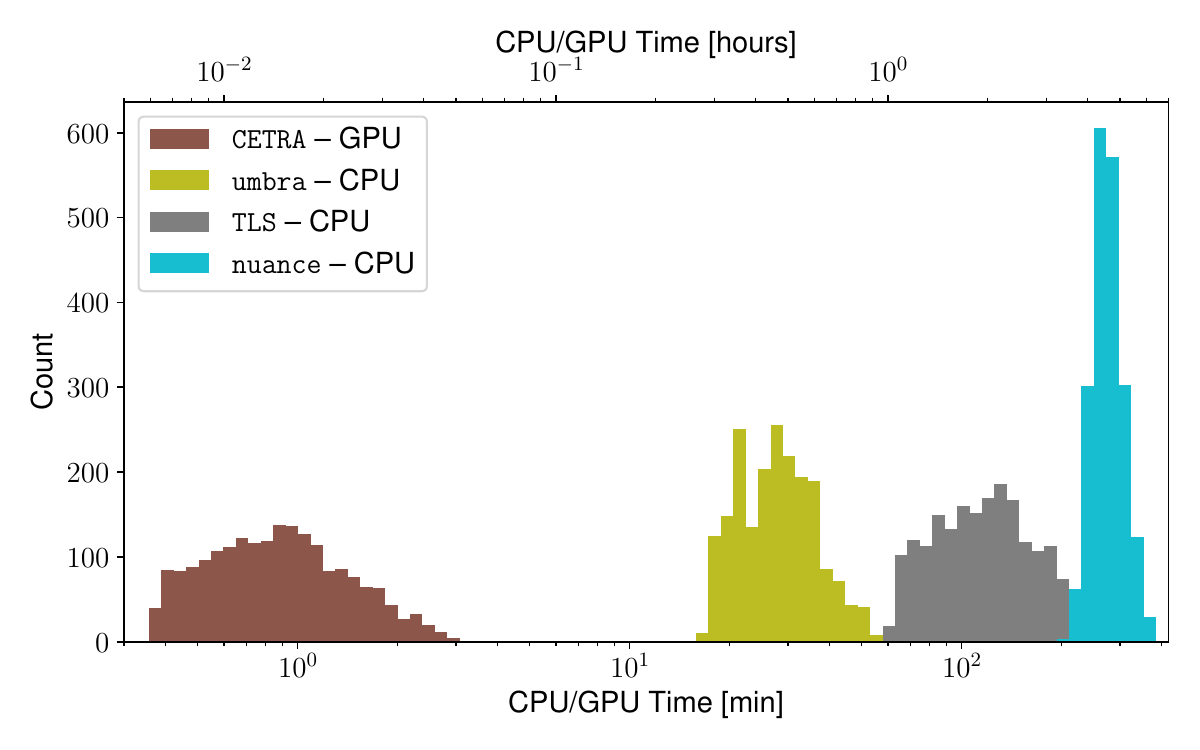}
    \caption{Computational performance of the different light curve filter and search algorithms. Left: runtime of the different filters on the $25$~s cadence PLATO data as a function of the window size used. Points indicate the mean computation time per light curve and the errors the 16th and 84th percentiles of the distribution. 
    Right: histograms of the per light curve CPU/GPU time of the different search algorithms for representative runs of each algorithm. CPU times from a search performed with {\tls} \citep{Hippke-2019-TLS} are included for comparison. {\nuance} used 10~min cadence data compared to 5~min for {\cetra}, {\umbra}, and {\tls}.}
    \label{fig:runtime}
\end{figure*}

\paragraph{Light curve filters} By default the light curve filters (biweight, Huber spline, Lowess, and YSD-Lowess) were run on the native 25~s cadence PLATO data. The majority were run on compute nodes with Intel Xeon 6242R @ 3.10 GHz CPUs, but some filter-window combinations were run on nodes with either Intel Xeon 8352Y @ 2.20 GHz or Intel Xeon 6226R @ 2.90 GHz CPUs. In all instances one CPU core was used per light curve. The time it took to run each filter is shown in the left panel of Fig.~\ref{fig:runtime}. Most methods increase their runtime with window-size as the window contains more points. The biweight and Lowess methods show similar performance, and YSD-Lowess takes slightly longer. Conversely the Huber spline gets faster with larger window-sizes as fewer knots are needed in the spline function. The Huber spline is also by far the fastest filter method, which is fortunate as it is the preferred one based on the tests presented in this work. Tests with the order of operations of binning and filtering (Sec.~\ref{sec:results-bin-filter}) suggest that binning the light curves before running the filters would not reduce the transit search performance, in which case all these runtimes are significantly reduced (see Table \ref{tab:overview_filter_bin}).

\paragraph{Search algorithms} {\cetra} was run on Nvidia H100 GPUs, while {\umbra} and {\nuance} were run on the same CPUs as the light curve filters. {\cetra} searches were performed using one GPU per light curve, {\umbra} searches used one CPU core per light curve, and {\nuance} used four CPU cores per light curve. The right panel of Fig.~\ref{fig:runtime} shows histograms of the CPU/GPU time it took to search each light curve for representative runs of the different search algorithms presented in this paper. For reference we also include the time it took to run {\tls} with comparable period and duration grids using a single core per light curve. As expected {\cetra} is by far the fastest code tested, taking $\sim 60$~s per light curve. Next is {\umbra} which takes $\sim30$~mins on average, followed by {\tls} which takes $\sim100$~mins. {\nuance} was run on lower cadence data and with a reduced duration grid compared to {\umbra}, yet it still takes by far the longest at $\sim 280$~mins. The issue with the duration limits employed throughout this paper (Sec.~\ref{sec:results-durations-limits}) means that all searches were run using too narrow a grid in durations. In tests with wider duration limits the {\cetra} runtime increased by 10-15\%, and similar or greater increases should be expected for the other codes. However, changes in the search grid are unlikely to change the order of the transit search algorithms from fastest to slowest.

\section{Summary and key recommendations}
\label{sec:conclusions}

\begin{figure*}[tbp!]
    \centering
    \includegraphics[width=1\linewidth]{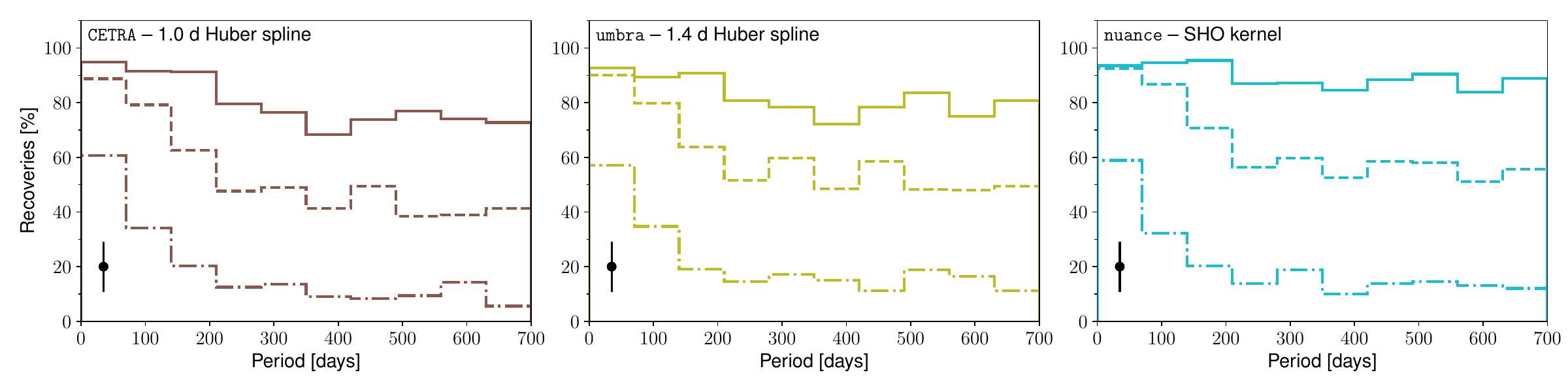}
    \caption{Percentage of transits as a function of orbital period across the joint simulations in three planet radius bins. {\cetra} and {\umbra} are shown for their respective optimal Huber spline filter windows. The black point shows the typical counting error in each bin. Though below the uncertainty in individual bins, overall improved recovery rates are visible when going from {\cetra} to {\umbra} to {\nuance}.}
    \label{fig:recovery_frac}
\end{figure*}

In this work we have performed a detailed performance comparison between different light curve filter and transit detection codes. The default PLATO transit search algorithm is {\cetra}, which requires a pre-filtered light curve for which we tested biweight, Huber spline, Lowess, and YSD-Lowess filters. We compare the {\cetra} performance to the WLS method of {\umbra} using the same filtered light curves; and the simultaneous filtering and detection of {\nuance}, which uses a GP to model the stellar variability at all proposed transit parameters. The simulated dataset used to perform these comparisons consists of 2000 stellar light curves simulated with with {\psls} and {\pyspot} for the nominal two year pointing of PLATO. Stellar parameters were taken from a subset of the PIC P1 + P2 sample, and the simulations include the effects of data-gaps, granulation, oscillations, star-spots and long-term systematics. Each light curve was used with two different transit signals: the inner orbits where $N_{\rm{tr}} \gtrsim 3$ and the outer orbits where $N_{\rm{tr}} \lesssim 3$, both with planet radii in the range $0.5-2.0~R_\oplus$. We do not consider multi-planet systems, or astrophysical false positives such as eclipsing binaries. 

Figure \ref{fig:recovery_frac} shows the recovery fractions as a function of orbital period and planet radius for each of the three transit search codes used in this study. {\cetra}, {\umbra}, and {\nuance} all shows comparable performance: at periods $P>150$~d they recover 70--90\% of planets with $1.5 \lesssim R_p/R_\oplus < 2.0$, 40--60\% of planets with $1.0 \lesssim R_p/R_\oplus < 1.5$, and 10--15\% of planets with $0.5 \lesssim R_p/R_\oplus < 1.0$. These numbers are the benchmark against which to evaluate the performance of new light curve filters or detection algorithms when tested on this simulated dataset. However, they should not be used to estimate PLATO planet yields in these regimes due to the limitations of the simulations. 

While the differences between {\cetra}, {\umbra}, and {\nuance} are small, they are nonetheless significant, as discussed in Sec.~\ref{sec:results}. This leads to our first set of recommendations. First, the warped transit shapes (Fig.~\ref{fig:warped_shapes}) used by {\umbra} are shown to be beneficial in detecting transits in pre-filtered light curves and should be implemented in {\cetra}, which is computationally much faster. Second, {\nuance} outperforms both {\cetra} and {\umbra} on hot, rapidly rotating, active stars (Fig.~\ref{fig:search_comparison_stellar}) and it is worth considering using {\nuance} in parallel to {\cetra} for this subset of stars, provided computational resources are available. Further testing would be needed, however, to optimize {\nuance}'s performance for these stars.

Regarding the choice of filter used with {\cetra} and {\umbra}, the results shown in Fig.~\ref{fig:recovery_frac} are for the Huber spline filter, which performed better than the biweight filter, and better than or similar to the Lowess and YSD-Lowess filters. We recommend the Huber spline for its speed, robustness against outliers and small data gaps, and overall performance in recovering signals (Fig.~\ref{fig:search_results}). We also recommend against using our implementation of the YSD-Lowess filter. While the YSD-Lowess filter performance can be comparable to the Huber spline, our method of interpolating over the peaks and troughs in the light curve results in inconsistent behaviour on the same feature in different parts of the light curve (see Fig.~\ref{fig:ysd_lowess_example}). Finally, while we ran our filters on the native 25~s PLATO cadence, we showed in Sec.~\ref{sec:results-bin-filter} that binning to 300 or even 600~s cadence before filtering does not significantly impact our ability to detect transits, but does reduce the computational cost of the Huber spline filter. 

The best window size to use with the Huber spline proved harder to pin down. The best performing window depends on the algorithm and the transit sample (inner/outer) considered, likely due to differing transit durations. Across the joint sample {\cetra} performs best with a 1.0~d window size and {\umbra} with a 1.4~d window size (Fig.~\ref{fig:search_results}). If WLS templates are implemented in {\cetra} the optimal window size may shift to match {\umbra}. Furthermore, we showed that (if resources are available) running multiple Huber spline windows results in an additional $4\%$ pt. signals recovered for two filters, and $6\%$ pt. for four filters in the joint sample (Fig.\ref{fig:filter_combos}), with window sizes ranging from $0.8-1.8$~d. Optimal filters can be selected from Table~\ref{tab:filter-combos} based on how many filters can be run and which transit sample is deemed most important.

The range of trial transit durations used with {\cetra}, {\umbra}, and {\nuance} was narrower than it should have been due to a misunderstanding with the use of the {\tls} package. We explored the impact of this issue in Sec.~\ref{sec:results-durations-limits}, but a full re-analysis of the light curves was beyond the scope of this work. Nevertheless, this problem has been addressed in the latest versions of {\cetra} and {\umbra} so that future work will not be affected. 

Finally, the new simulated dataset used in this work was generated with our best understanding of stellar variability signals present in PLATO, and PLATO's observational characteristics. However, the volume of simulated sources required necessitated the use of a simplified PLATO systematics model with {\psls}. More detailed modelling of systematics is possible through pixel level simulations, but real PLATO data is likely to contain effects our simulators do not currently account for. We recommend that the performance of the filters, filter windows, and transit search methods presented here should be re-evaluated on real PLATO data using injection-recovery tests once data from a few observational quarters are available.

Future work should focus on implementing WLS in {\cetra} and benchmarking its performance to {\umbra} using the updated duration limits. There is also scope for improvements to the transit templates used: adding to the range of transit shapes, e.g. adding templates with high impact parameters; and implementing a wider range of smoothing options for generating WLS templates. Finally, more work is also needed to determine if {\nuance} is worth the additional computational cost for a subset of the stellar sample.

\begin{acknowledgements}
This work presents results from the European Space Agency (ESA) space mission
PLATO. The PLATO payload, the PLATO Ground Segment and PLATO data processing
are joint developments of ESA and the PLATO Mission Consortium (PMC). Funding for
the PMC is provided at national levels, in particular by countries participating in the
PLATO Multilateral Agreement (Austria, Belgium, Czech Republic, Denmark, France,
Germany, Italy, Netherlands, Portugal, Spain, Sweden, Switzerland, Norway, and United
Kingdom) and institutions from Brazil. Members of the PLATO Consortium can be found
at \url{https://platomission.com/}. The ESA PLATO mission website is
\url{https://www.cosmos.esa.int/plato}. We thank the teams working for PLATO for all their work. This work was supported by the UK Space Agency as part of its support for the UK component of the PLATO Mission. SA also acknowledges support from the ERC via CoG 865624. LCS is supported by UKRI-STFC grants ST/X001628/1 and ST/X001571/1. LM acknowledge support from PLATO ASI-INAF agreements n. 2022-28-HH.0. The authors would like to acknowledge contributions made by Eric Feigelson and Marco Montalto during the preparation of this manuscript. We also thank the referee Michael Hippke and the editor José A. Caballero for their feedback.
\end{acknowledgements}

\bibliographystyle{aa}
\bibliography{filtersearch.bib}

\begin{appendix}
\onecolumn

\section{Tables of results}
\label{app:results-table}

Table~\ref{tab:overview} lists the results of transit searches performed with {\cetra}, {\umbra}, and {\nuance}. The results of combining {\cetra} and {\umbra} search results obtained with different filters and window sizes are listed in Tab.~\ref{tab:filter-combos}.

\begin{table*}[bp!]
    \centering
    \caption{Number of correctly recovered signals for all tested light curve filters and transit search codes.}
    \label{tab:overview}
    \begin{tabular}{llrrlrrr}
        \hline
        \hline
         Light curve & Filter & Window & Bin Size & Search & inner & outer & joint \\
         & & [d] & [s] & & 2000 & 2000 & 4000 \\
         \hline
         Simple & ... & ... & 300 & {\cetra} & 1570 & 1476 & 3046 \\
         Simple & ... & ... & 300 & {\umbra}\tablefootmark{a} & 1553 & 1457 & 3010 \\
         Simple & ... & ... & 600 & \nuance & 1604 & 1527 & 3131 \\
         \hline
         Full & SHO Kernel & ... & 600 & \nuance & 1373 & 1076 & 2449 \\
         \hline
         Full & Biweight & 0.7 & 300 & {\cetra} & ... & 625 & ... \\
         Full & Biweight & 1.0 & 300 & {\cetra} & ... & 809 & ... \\
         Full & Biweight & 1.4 & 300 & {\cetra} & ... & 845 & ... \\
         Full & Biweight & 2.0 & 300 & {\cetra} & ... & 797 & ... \\
         Full & Biweight & 0.7 & 300 & {\umbra} & ... & 775 & ... \\
         Full & Biweight & 1.0 & 300 & {\umbra} & ... & 916 & ... \\
         Full & Biweight & 1.4 & 300 & {\umbra} & ... & 975 & ... \\
         Full & Biweight & 2.0 & 300 & {\umbra} & ... & 915 & ... \\
         \hline 
         Full & Huber spline & 0.6 & 300 & {\cetra} & 1164 &  531 & 1695 \\
         Full & Huber spline & 0.8 & 300 & {\cetra} & 1282 &  766 & 2048 \\
         Full & Huber spline & 1.0 & 300 & {\cetra} & 1309 &  889 & 2198 \\
         Full & Huber spline & 1.2 & 300 & {\cetra} & 1276 &  902 & 2178 \\
         Full & Huber spline & 1.4 & 300 & {\cetra} & 1219 & 922 & 2141 \\
         Full & Huber spline & 1.6 & 300 & {\cetra} & 1177 &  897 & 2074 \\
         Full & Huber spline & 1.8 & 300 & {\cetra} & 1112 &  857 & 1969 \\
         Full & Huber spline & 2.0 & 300 & {\cetra} & 1060 &  814 & 1874 \\
         Full & Huber spline & 0.6 & 300 & {\umbra} & 1187 &  704 & 1891 \\
         Full & Huber spline & 0.8 & 300 & {\umbra} & 1288 &  879 & 2167 \\
         Full & Huber spline & 1.0 & 300 & {\umbra} & 1324 &  945 & 2269 \\
         Full & Huber spline & 1.2 & 300 & {\umbra} & 1325 &  979 & 2304 \\
         Full & Huber spline & 1.4 & 300 & {\umbra} & 1299 & 1009 & 2308 \\
         Full & Huber spline & 1.6 & 300 & {\umbra} & 1278 &  986 & 2264 \\
         Full & Huber spline & 1.8 & 300 & {\umbra} & 1241 &  978 & 2219 \\
         Full & Huber spline & 2.0 & 300 & {\umbra} & 1199 &  944 & 2143 \\
         \hline
         Full & Lowess & 1.0 & 300 & {\cetra} & ... & 631 & ... \\
         Full & Lowess & 1.4 & 300 & {\cetra} & ... & 796 & ... \\
         Full & Lowess & 2.0 & 300 & {\cetra} & ... & 847 & ... \\
         Full & Lowess & 2.4 & 300 & {\cetra} & ... & 824 & ... \\
         Full & Lowess & 1.0 & 300 & {\umbra}\tablefootmark{b} & ... & 783 & ... \\
         Full & Lowess & 1.4 & 300 & {\umbra}\tablefootmark{b} & ... & 926 & ... \\
         Full & Lowess & 2.0 & 300 & {\umbra}\tablefootmark{b} & ... & 1033 & ... \\
         Full & Lowess & 2.4 & 300 & {\umbra}\tablefootmark{b} & ... & 1025 & ... \\
         \hline
         Full & YSD-Lowess & 1.0 & 300 & {\cetra} & ... & 531 & ... \\
         Full & YSD-Lowess & 1.4 & 300 & {\cetra} & ... & 735 & ... \\
         Full & YSD-Lowess & 2.0 & 300 & {\cetra} & ... & 854 & ... \\
         Full & YSD-Lowess & 2.4 & 300 & {\cetra} & ... & 852 & ... \\
         Full & YSD-Lowess & 1.0 & 300 & {\umbra}\tablefootmark{b} & ... & 704 & ... \\
         Full & YSD-Lowess & 1.4 & 300 & {\umbra}\tablefootmark{b} & ... & 836 & ... \\
         Full & YSD-Lowess & 2.0 & 300 & {\umbra}\tablefootmark{b} & ... & 932 & ... \\
         Full & YSD-Lowess & 2.4 & 300 & {\umbra}\tablefootmark{b} & ... & 931 & ... \\
         \hline
    \end{tabular}
    \tablefoot{These results are visualized in Fig.~\ref{fig:search_results}. Not all filters and codes were tested on the inner sample. 
    \tablefoottext{a}{Since the simple light curves were variability free and did not need filtering {\umbra} was run with TLS templates.}
    \tablefoottext{b}{Because the Lowess and YSD-Lowess methods use a tricube weighting function across the filter window, {\umbra}'s WLS templates were computed with the same tricube weighting for these filters.}}
\end{table*}

\begin{table*}[tbp!]
    \centering
    \caption{Correctly recovered signals and preferred light curve filters and window sizes when combining the results from different filter-window combinations.}
    \label{tab:filter-combos}
    \resizebox{\linewidth}{!}{
    \begin{tabular}{lllrrcccccccccc}
        \hline
        \hline
        Search & Sample & Filters & No. Filters & Recoveries & \multicolumn{8}{c}{Huber spline} & Lowess & YSD-Lowess \\
         &  &  &  &  & 0.6~d & 0.8~d & 1.0~d & 1.2~d & 1.4~d & 1.6~d & 1.8~d & 2.0~d & 2.0~d & 2.0~d \\ 
        \hline
        {\cetra} & outer & Any & 1 &  922 & ... & ... & ... & ... & \checkmark & ... & ... & ... & ... & ... \\ 
        {\cetra} & outer & Any & 2 & 1011 & ... & ... & \checkmark & ... & \checkmark & ... & ... & ... & ... & ... \\ 
        {\cetra} & outer & Any & 3 & 1069 & ... & ... & \checkmark & ... & ... & \checkmark & ... & ... & ... & \checkmark \\ 
        {\cetra} & outer & Any & 4 & 1100 & ... & ... & \checkmark & ... & \checkmark & ... & ... & \checkmark & ... & \checkmark \\ 
        {\cetra} & outer & Any & 5 & 1123 & ... & ... & \checkmark & \checkmark & \checkmark & ... & \checkmark & ... & ... & \checkmark \\ 
        {\cetra} & outer & Any & 6 & 1138 & ... & \checkmark & \checkmark & \checkmark & \checkmark & ... & \checkmark & ... & ... & \checkmark \\ 
        \hline
        {\cetra} & outer & Huber spline & 1 &  922 & ... & ... & ... & ... & \checkmark & ... & ... & ... & ... & ... \\ 
        {\cetra} & outer & Huber spline & 2 & 1011 & ... & ... & \checkmark & ... & \checkmark & ... & ... & ... & ... & ... \\ 
        {\cetra} & outer & Huber spline & 3 & 1054 & ... & ... & \checkmark & ... & \checkmark & ... & ... & \checkmark & ... & ... \\ 
        {\cetra} & outer & Huber spline & 4 & 1084 & ... & ... & \checkmark & \checkmark & \checkmark & ... & \checkmark & ... & ... & ... \\ 
        {\cetra} & outer & Huber spline & 5 & 1107 & ... & \checkmark & \checkmark & \checkmark & \checkmark & ... & \checkmark & ... & ... & ... \\ 
        {\cetra} & outer & Huber spline & 6 & 1121 & ... & \checkmark & \checkmark & \checkmark & \checkmark & \checkmark & \checkmark & ... & ... & ... \\
        \hline
        {\cetra} & inner & Huber spline & 1 & 1309 & ... & ... & \checkmark & ... & ... & ... & ... & ... & ... & ... \\ 
        {\cetra} & inner & Huber spline & 2 & 1349 & ... & \checkmark & ... & \checkmark & ... & ... & ... & ... & ... & ... \\ 
        {\cetra} & inner & Huber spline & 3 & 1367 & ... & \checkmark & \checkmark & ... & ... & \checkmark & ... & ... & ... & ... \\ 
        {\cetra} & inner & Huber spline & 4 & 1381 & \checkmark & \checkmark & \checkmark & ... & ... & \checkmark & ... & ... & ... & ... \\ 
        {\cetra} & inner & Huber spline & 5 & 1386 & \checkmark & \checkmark & \checkmark & \checkmark & ... & \checkmark & ... & ... & ... & ... \\ 
        {\cetra} & inner & Huber spline & 6 & 1392 & \checkmark & \checkmark & \checkmark & \checkmark & \checkmark & ... & \checkmark & ... & ... & ... \\
        \hline
        {\cetra} & joint & Huber spline & 1 & 2198 & ... & ... & \checkmark & ... & ... & ... & ... & ... & ... & ... \\ 
        {\cetra} & joint & Huber spline & 2 & 2345 & ... & ... & \checkmark & ... & ... & \checkmark & ... & ... & ... & ... \\ 
        {\cetra} & joint & Huber spline & 3 & 2404 & ... & \checkmark & \checkmark & ... & ... & \checkmark & ... & ... & ... & ... \\ 
        {\cetra} & joint & Huber spline & 4 & 2451 & ... & \checkmark & \checkmark & ... & \checkmark & ... & \checkmark & ... & ... & ... \\ 
        {\cetra} & joint & Huber spline & 5 & 2486 & ... & \checkmark & \checkmark & \checkmark & \checkmark & ... & \checkmark & ... & ... & ... \\ 
        {\cetra} & joint & Huber spline & 6 & 2503 & \checkmark & \checkmark & \checkmark & \checkmark & \checkmark & ... & \checkmark & ... & ... & ... \\
        \hline
        {\umbra} & outer & Any & 1 & 1033 & ... & ... & ... & ... & ... & ... & ... & ... & \checkmark & ... \\ 
        {\umbra} & outer & Any & 2 & 1109 & ... & ... & ... & ... & ... & ... & \checkmark & ... & \checkmark & ... \\ 
        {\umbra} & outer & Any & 3 & 1148 & ... & ... & \checkmark & ... & ... & ... & \checkmark & ... & ... & \checkmark \\ 
        {\umbra} & outer & Any & 4 & 1174 & ... & \checkmark & \checkmark & ... & ... & ... & \checkmark & ... & ... & \checkmark \\ 
        {\umbra} & outer & Any & 5 & 1195 & ... & \checkmark & \checkmark & ... & ... & ... & \checkmark & \checkmark & ... & \checkmark \\ 
        {\umbra} & outer & Any & 6 & 1209 & ... & \checkmark & \checkmark & ... & ... & \checkmark & \checkmark & \checkmark & ... & \checkmark \\ 
        \hline
        {\umbra} & outer & Huber spline & 1 & 1009 & ... & ... & ... & ... & \checkmark & ... & ... & ... & ... & ... \\
        {\umbra} & outer & Huber spline & 2 & 1093 & ... & ... & \checkmark & ... & ... & ... & \checkmark & ... & ... & ... \\
        {\umbra} & outer & Huber spline & 3 & 1135 & ... & \checkmark & ... & ... & \checkmark & ... & \checkmark & ... & ... & ... \\
        {\umbra} & outer & Huber spline & 4 & 1160 & ... & \checkmark & \checkmark & ... & \checkmark & ... & \checkmark & ... & ... & ... \\
        {\umbra} & outer & Huber spline & 5 & 1180 & ... & \checkmark & \checkmark & ... & \checkmark & ... & \checkmark & \checkmark & ... & ... \\
        {\umbra} & outer & Huber spline & 6 & 1193 & ... & \checkmark & \checkmark & ... & \checkmark & \checkmark & \checkmark & \checkmark & ... & ... \\
        \hline
        {\umbra} & inner & Huber spline & 1 & 1325 & ... & ... & ... & \checkmark & ... & ... & ... & ... & ... & ... \\
        {\umbra} & inner & Huber spline & 2 & 1381 & ... & \checkmark & ... & ... & ... & \checkmark & ... & ... & ... & ... \\
        {\umbra} & inner & Huber spline & 3 & 1409 & ... & \checkmark & ... & \checkmark & ... & \checkmark & ... & ... & ... & ... \\
        {\umbra} & inner & Huber spline & 4 & 1417 & \checkmark & \checkmark & ... & \checkmark & ... & \checkmark & ... & ... & ... & ... \\
        {\umbra} & inner & Huber spline & 5 & 1425 & \checkmark & \checkmark & \checkmark & \checkmark & ... & \checkmark & ... & ... & ... & ... \\
        {\umbra} & inner & Huber spline & 6 & 1432 & \checkmark & \checkmark & \checkmark & \checkmark & ... & \checkmark & ... & \checkmark & ... & ... \\
        \hline
        {\umbra} & joint & Huber spline & 1 & 2308 & ... & ... & ... & ... & \checkmark & ... & ... & ... & ... & ... \\
        {\umbra} & joint & Huber spline & 2 & 2467 & ... & \checkmark & ... & ... & ... & ... & \checkmark & ... & ... & ... \\
        {\umbra} & joint & Huber spline & 3 & 2537 & ... & \checkmark & ... & \checkmark & ... & ... & \checkmark & ... & ... & ... \\
        {\umbra} & joint & Huber spline & 4 & 2573 & ... & \checkmark & \checkmark & ... & \checkmark & ... & \checkmark & ... & ... & ... \\
        {\umbra} & joint & Huber spline & 5 & 2602 & ... & \checkmark & \checkmark & ... & \checkmark & ... & \checkmark & \checkmark & ... & ... \\
        {\umbra} & joint & Huber spline & 6 & 2618 & \checkmark & \checkmark & \checkmark & ... & \checkmark & ... & \checkmark & \checkmark & ... & ... \\
        \hline
    \end{tabular}
    }
    \tablefoot{These results are visualized in Fig.~\ref{fig:filter_combos}.}
\end{table*}

\section{Note on thresholds}
\label{app:thresholds}

Our analysis of the performance of the different filters and search codes was performed without applying any kind of detection threshold, either on the periodogram peak values or another detection significance statistic. It is reasonable to expect that applying thresholds and comparing on recovered true positives (matching period/epoch, above threshold) our analysis might have been qualitatively and quantitatively different. We perform here a simple analysis, where we apply permissive thresholds to the {\cetra}, {\umbra}, and {\nuance} results, to argue that our qualitative conclusions likely hold.

\begin{figure*}
    \centering
    \includegraphics[width=1\linewidth]{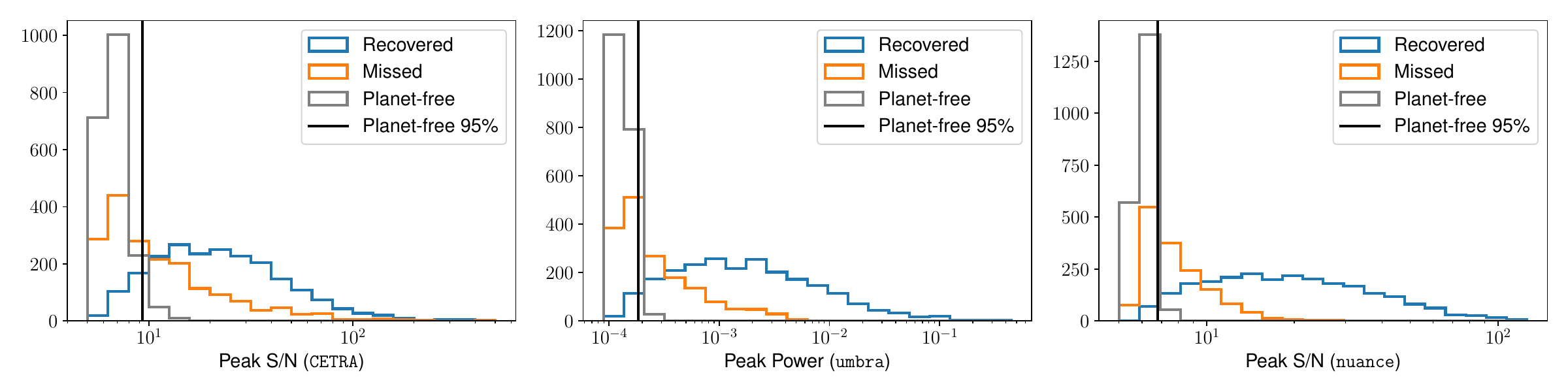}
    \caption{Comparison of periodogram peak strength in the simple planet-free light curves (grey) to peak strength in the joint sample split by recovered (blue) and missed (orange) signals for {\cetra} (left), {\umbra} (middle), and {\nuance} (right). The vertical black line marks the 95th percentile of the simple planet-free distribution.}
    \label{fig:thresholds}
\end{figure*}

Figure~\ref{fig:thresholds} shows peak S/N (power) distributions for {\cetra}, {\umbra}, and {\nuance}. The simple planet-free peak power distributions are representative of the peak power distributions obtained from planet-free perfectly filtered light curve and the 95th percentile of this distribution represents a permissive threshold. A permissive metric is desirable to achieve PLATO's aim of finding Earth-like planets in the habitable zone, which will show only a few transits in the two year nominal baseline. Further pipeline steps will be used to vet the initial list of candidate signals and prioritize targets for follow-up. Using this threshold produces few False Negatives (recovered signals with peaks below the threshold), and many False Positives (missed signals with peaks above the threshold). For all search codes the planet-free and recovered histograms are roughly equally well-separated, and we can expect no qualitative impact on our results from applying thresholds. 

\section{Supplementary figures}
\label{app:supplementary-figures}

\begin{figure*}[tbp!]
    \sidecaption
    \includegraphics[width=12cm]{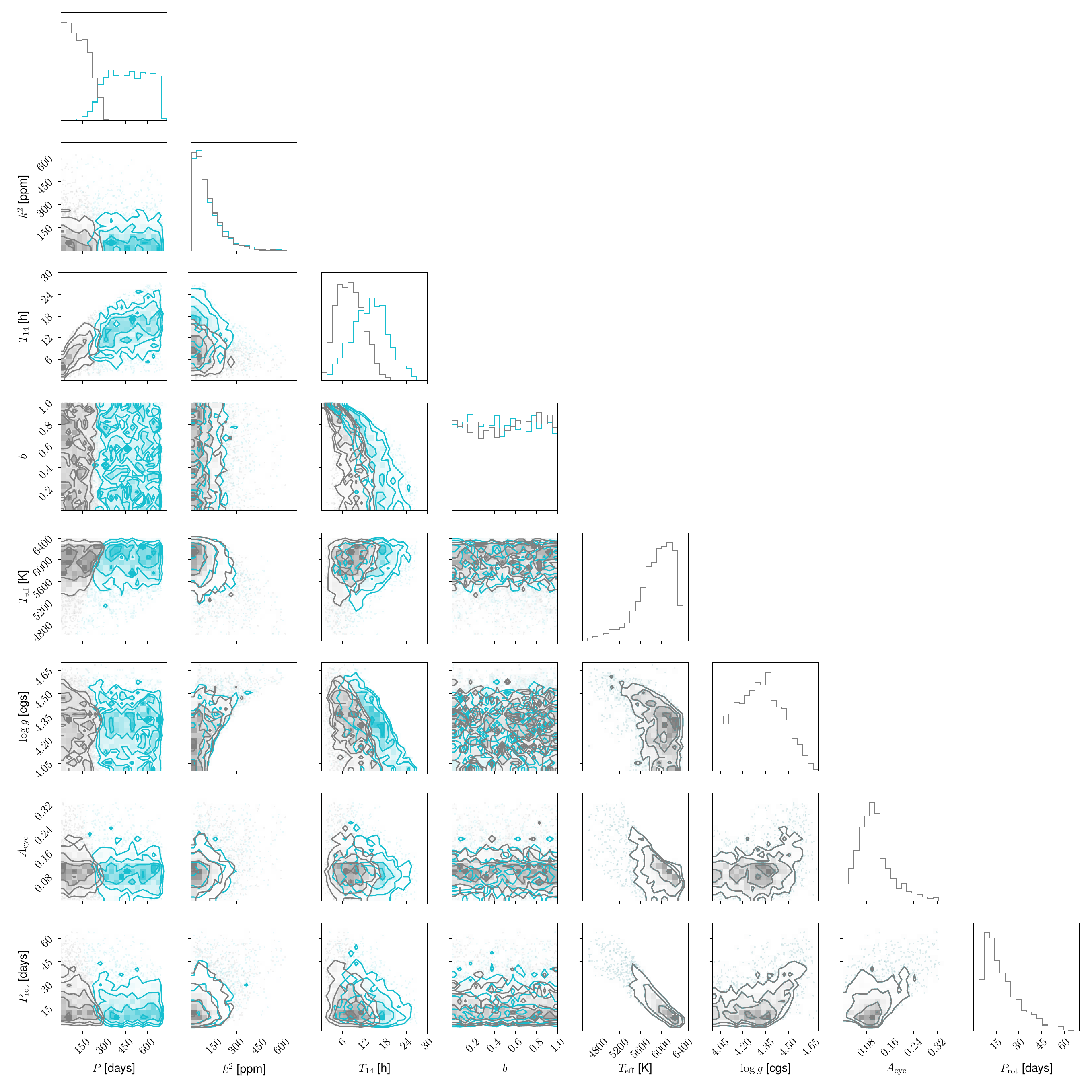}
    \caption{Correlations between transit and stellar parameters for the inner (grey) and outer (cyan) samples.}
    \label{fig:correlations}
\end{figure*}

\begin{figure*}[tbp!]
    \sidecaption
    \includegraphics[width=12cm]{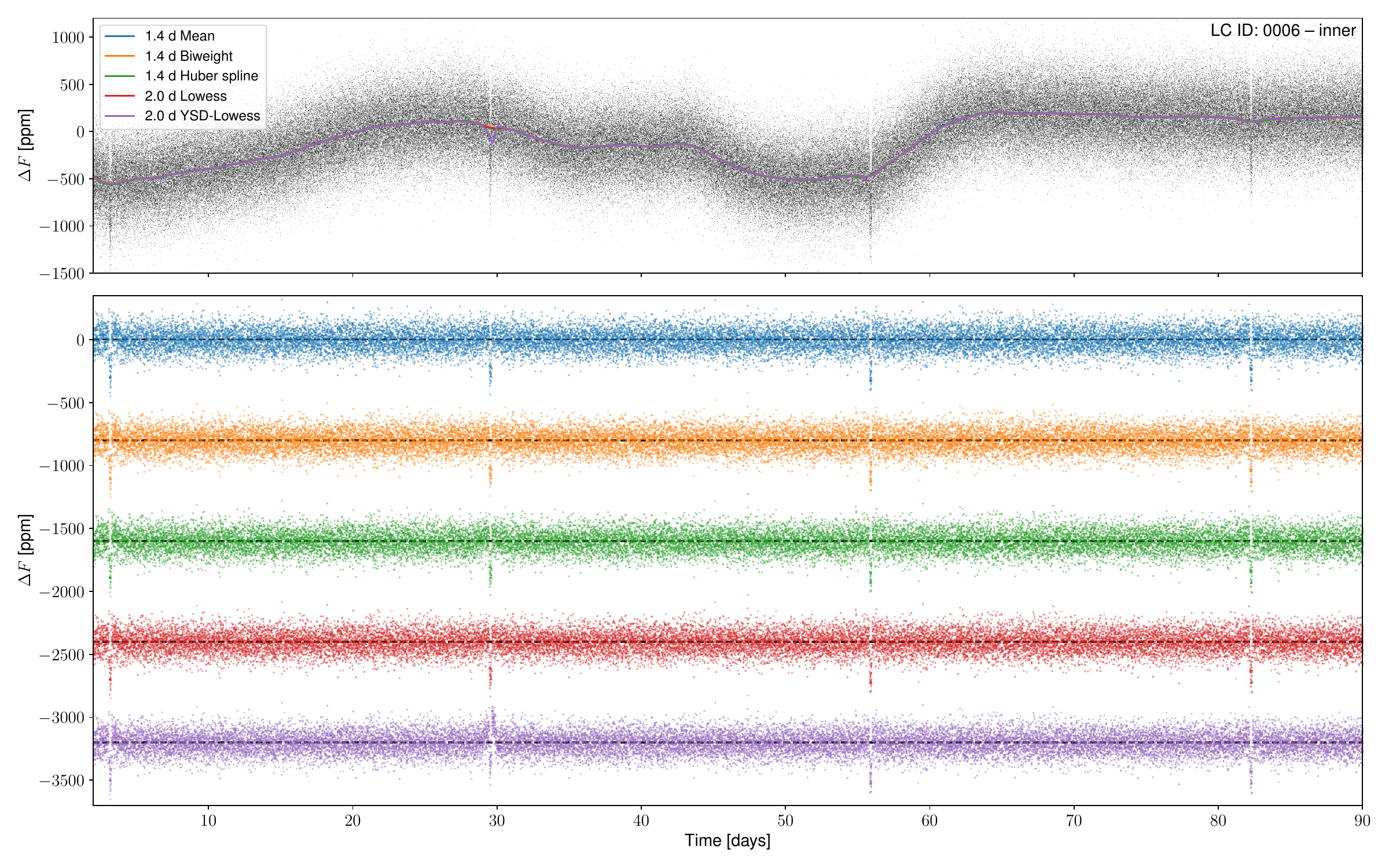}
    \caption{Same as Fig.~\ref{fig:filters_quarter} but for a quiet star.}
    \label{fig:filters_quiet}
    \sidecaption
    \includegraphics[width=12cm]{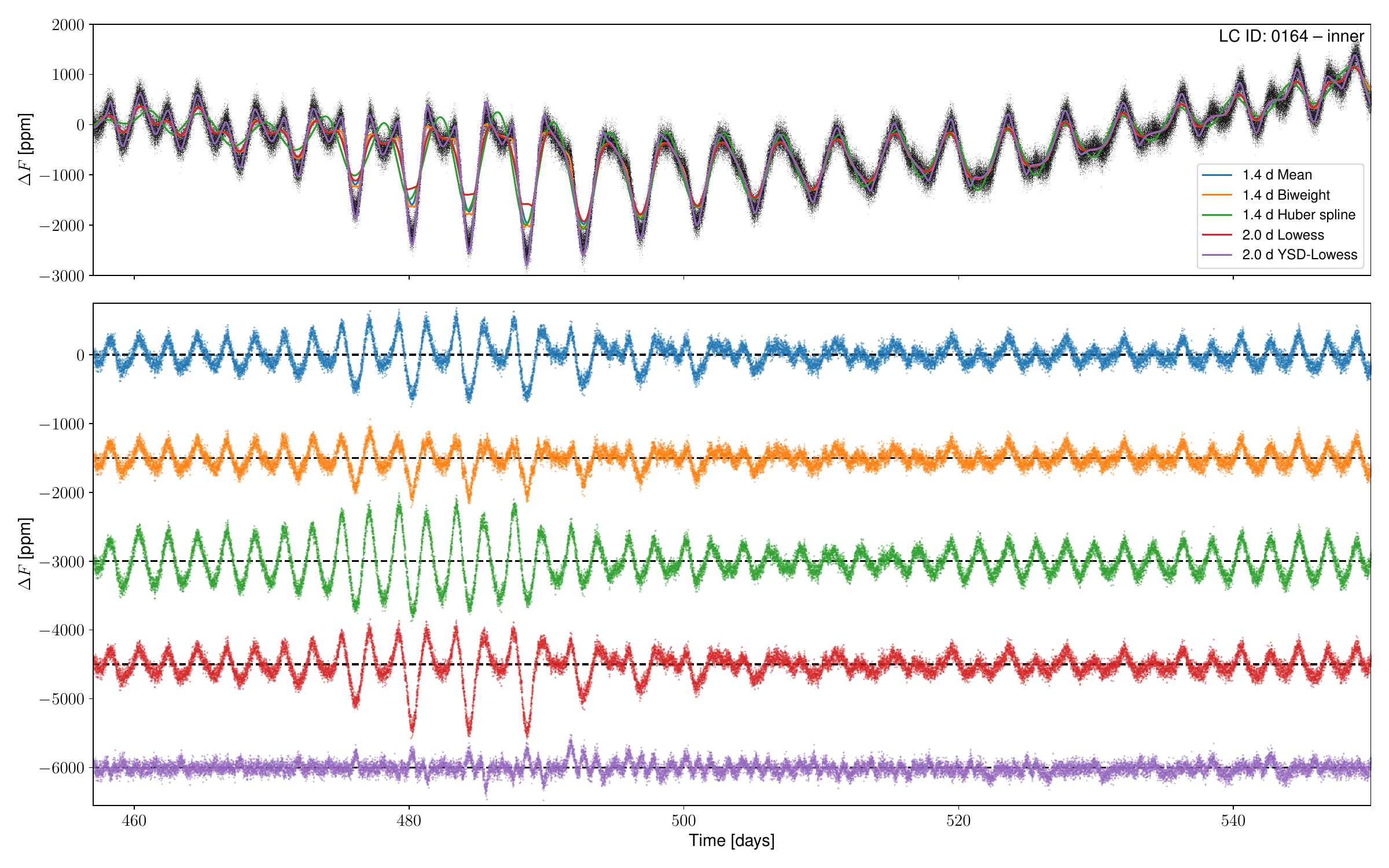}
    \caption{Same as Fig.~\ref{fig:filters_quarter} but for a highly active star.}
    \label{fig:filters_active}
\end{figure*}

\begin{figure*}[tbp!]
    \sidecaption
    \includegraphics[width=12cm]{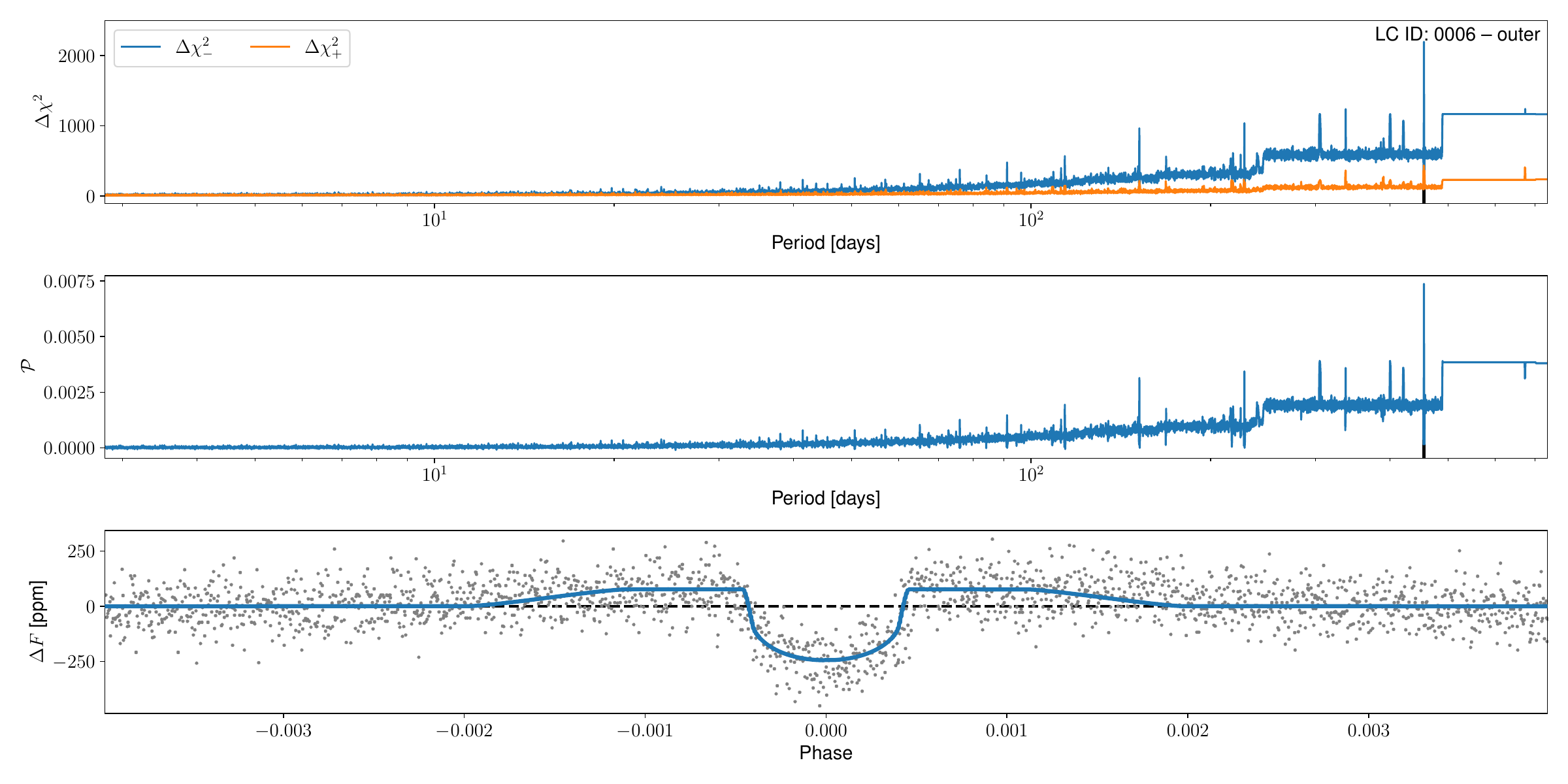}
    \caption{Example of the {\umbra} periodogram using WLS transit templates, obtained on a 1.4~d Huber spline filtered light curve. Top: the $\Delta \chi^2$ values computed for this light curve. Middle: The periodogram power ($\mathcal{P}$) computed from the $\Delta \chi^2$ values. Bottom: the light curve phase-folded to the best period and overlaid with the best-fit WLS transit template.}
    \label{fig:example_umbra_wls}
\end{figure*}

\begin{figure*}[tbp!]
    \sidecaption
    \includegraphics[width=12cm]{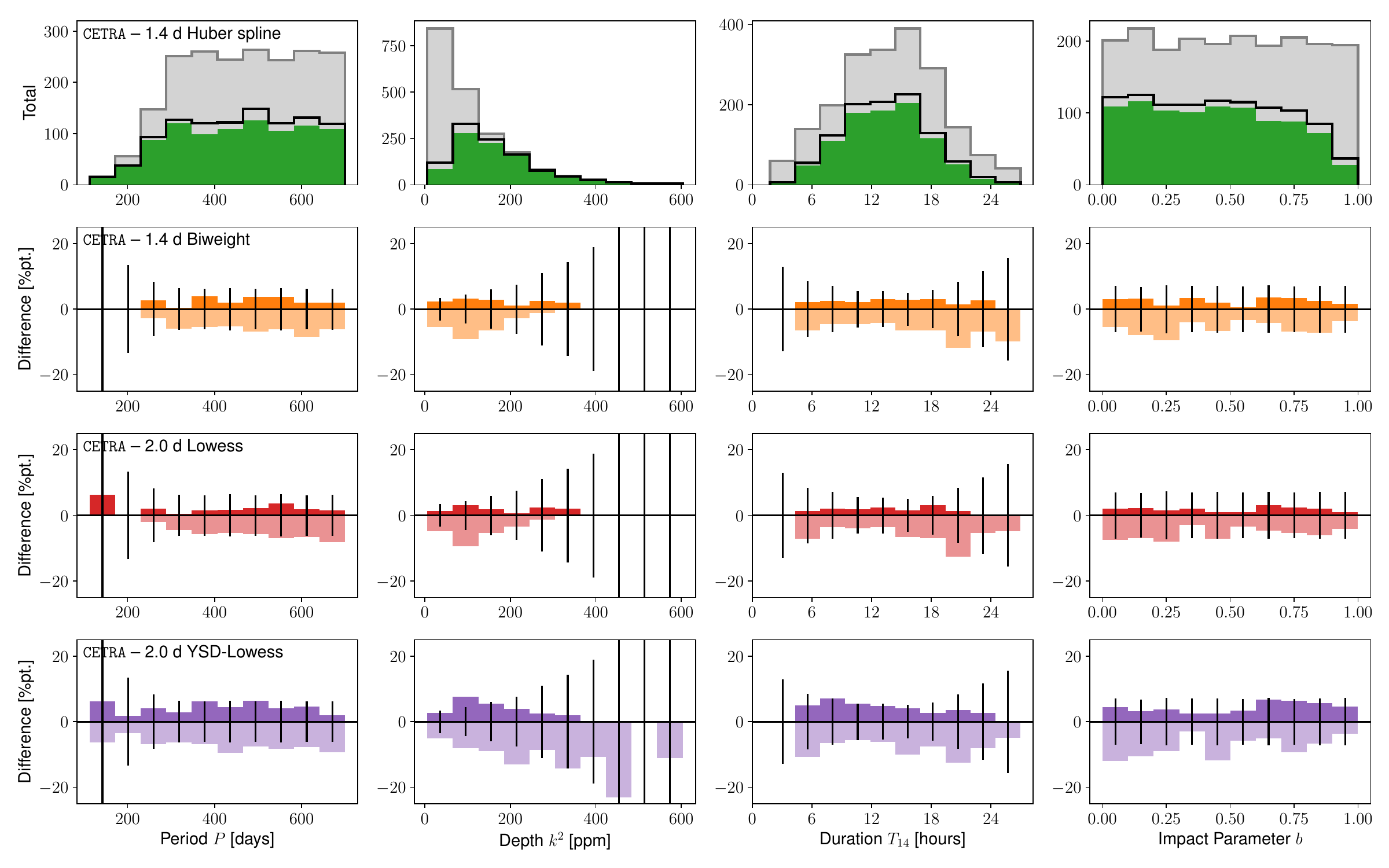}
    \caption{Same as Fig.~\ref{fig:search_comparison_transit}, but comparing light curve filters searched with {\cetra}.}
    \label{fig:filter_comparison_transit}
    \sidecaption
    \includegraphics[width=12cm]{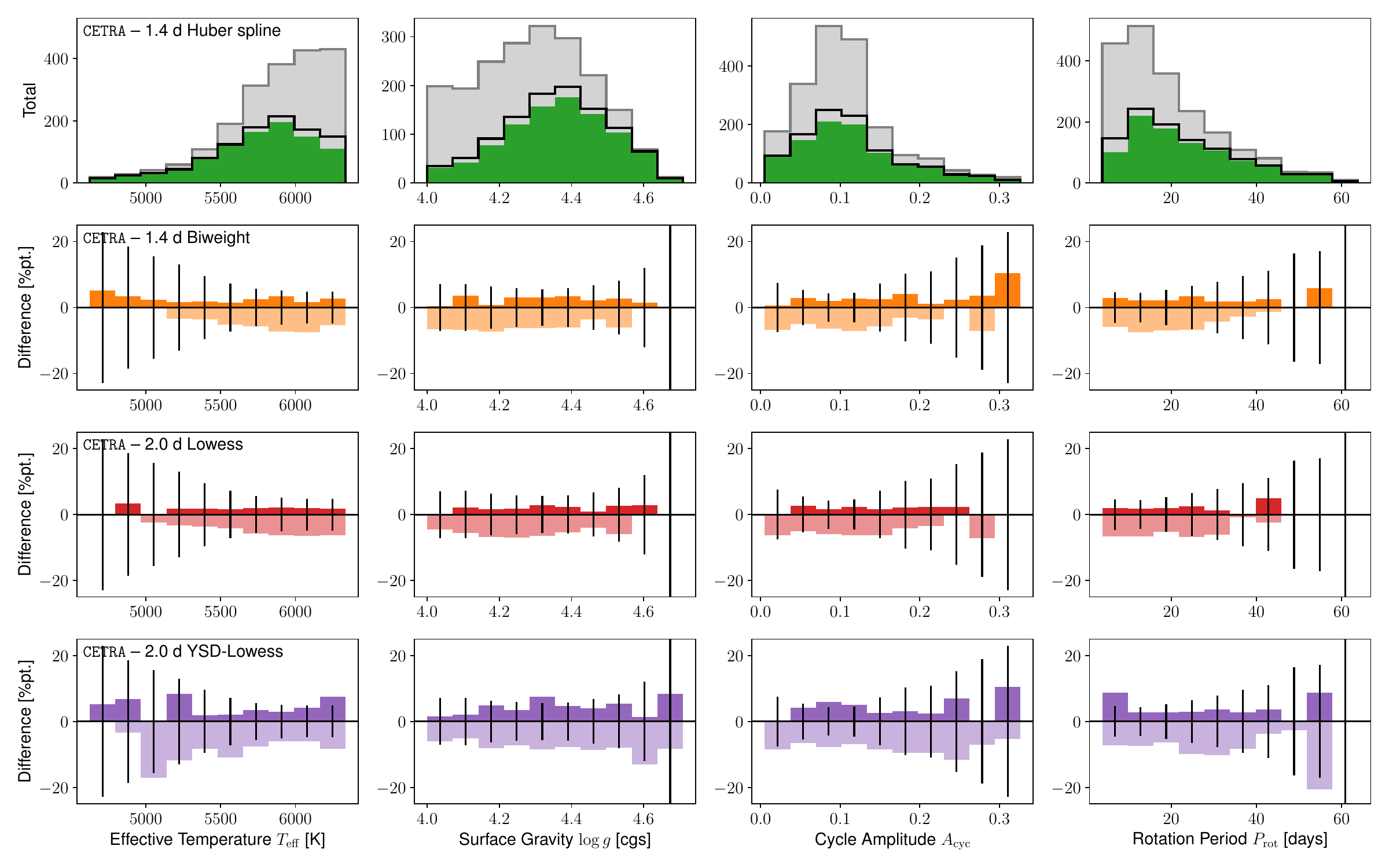}
    \caption{Same as Fig.~\ref{fig:search_comparison_stellar}, but comparing light curve filters searched with {\cetra}.}
    \label{fig:filter_comparison_stellar}
\end{figure*}

Figure~\ref{fig:correlations} shows the full correlations between key stellar and transit parameters for the inner and outer transit sample. There is a strong anti-correlation between $T_{\rm{eff}}$ and $P_{\rm{rot}}$, which contributes to the lower detection rate for shallow, long-period transits around the hotter stars. The weaker anti-correlation between $T_{\rm{eff}}$ and the cycle amplitude partly compensates for this effect. There is also a well-known correlation between shorter transits $T_{14}$ and higher impact parameters $b$. The duration grids in this work neglected this effect (see Sec.~\ref{sec:results-durations-limits}), which likely accounts for the decrease in performance at high impact parameters seen in {\cetra} and {\umbra}. {\nuance} performs better at high $b$ only because it implements no period dependent limits on the allowed durations, though this can also cause increased noise in the periodogram.
Figures~\ref{fig:filters_quiet} and \ref{fig:filters_active} accompany Fig.~\ref{fig:filters_quarter}. They show additional examples of the effect of light curve filters on stars across a range of activity levels.
Figure~\ref{fig:example_umbra_wls} shows example output obtained from running {\umbra}'s WLS on one of the simulated PLATO light curves described in this paper. The {\umbra} definition of periodogram power removes some of the trend in the $\Delta\chi^2$ with period, but preserves those parts of it which are due to signal. The WLS template provides a good match to the filtered transit shape.
Figures~\ref{fig:filter_comparison_transit} and \ref{fig:filter_comparison_stellar} compare the performance of the different filter algorithms at comparable window sizes when searched with {\cetra}. The biweight, Lowess, and YSD-Lowess filters recover some signals that are missed by the Huber spline filter, though this difference is only marginally significant for the YSD-Lowess filter. The YSD-Lowess filter performs slightly better on hot, rapidly rotating, stars which makes sense as it was designed to handle young, rapidly rotating stars. Conversely all filters miss a significant number of signals that are recovered by the Huber spline.

\end{appendix}

\end{document}